\documentclass[nofootinbib,a4paper,aps,prA,reprint,twocolumn,preprintnumbers,amsmath,amssymb,10pt, superscriptaddress]{revtex4-1}
\usepackage{verbatim}
\usepackage{graphicx}
\usepackage{color}
\usepackage{tikz}
\usepackage[colorlinks=true,citecolor=blue,linkcolor=blue,urlcolor=blue]{hyperref}
\usepackage{braket}
\usepackage[T1]{fontenc}
\usepackage[normalem]{ulem}
\usepackage{upgreek}
\usepackage[percent]{overpic}
\usepackage{graphicx,xcolor}% Include figure files
\usepackage{dcolumn}% Align table columns on decimal point
\usepackage{bm}% bold math
\usepackage[shortlabels]{enumitem}
\usepackage{pifont}
\usepackage[utf8]{inputenc}

\usepackage{graphicx,xcolor}% Include figure files
\usepackage{dcolumn}% Align table columns on decimal point
\usepackage{bm}% bold math
\usepackage[percent]{overpic}
\usepackage{physics}
\usepackage{mathtools}
\usepackage{algorithm}
\usepackage{algpseudocode}

\definecolor{darkGreen}{RGB}{0,110,0}
\definecolor{darkBlue}{RGB}{0,0,130}

\usepackage{ulem} % cancel text

\newcommand{\be}{\begin{equation}}
\newcommand{\ee}{\end{equation}}

\newcommand{\Z}{\mathbb{Z}}

\newcommand{\multiline}[1]{%
  \begin{tabularx}{\dimexpr\linewidth-\ALG@thistlm}[t]{@{}X@{}}
    #1
  \end{tabularx}
}

\usepackage{bm}% bold math

\begin{document}

\title{Many-body magic via Pauli-Markov chains - from criticality to gauge theories}

\author{Poetri Sonya Tarabunga}
\affiliation{The Abdus Salam International Centre for Theoretical Physics (ICTP), Strada Costiera 11, 34151 Trieste, Italy}
\affiliation{SISSA, Via Bonomea 265, 34136 Trieste, Italy}
\affiliation{INFN, Sezione di Trieste, Via Valerio 2, 34127 Trieste, Italy}

\author{Emanuele Tirrito}
\affiliation{The Abdus Salam International Centre for Theoretical Physics (ICTP), Strada Costiera 11, 34151 Trieste, Italy}
\affiliation{Pitaevskii BEC Center, CNR-INO and Dipartimento di Fisica, Università di Trento, Via Sommarive 14, Trento, I-38123, Italy}
\author{Titas Chanda}
\affiliation{The Abdus Salam International Centre for Theoretical Physics (ICTP), Strada Costiera 11, 34151 Trieste, Italy}
\affiliation{Department of Physics, Indian Institute of Technology Indore, Khandwa Road, Simrol, Indore 453552, India}
\author{Marcello Dalmonte}
\affiliation{The Abdus Salam International Centre for Theoretical Physics (ICTP), Strada Costiera 11, 34151 Trieste, Italy}
\affiliation{SISSA, Via Bonomea 265, 34136 Trieste, Italy}

\begin{abstract}
We introduce a method to measure many-body magic in quantum systems based on a statistical exploration of Pauli strings via Markov chains. We demonstrate that sampling such Pauli-Markov chains gives ample flexibility in terms of partitions where to sample from: in particular, it enables to efficiently extract the magic contained in the correlations between widely-separated subsystems, which characterizes the nonlocality of magic. Our method can be implemented in a variety of situations. We describe an efficient sampling procedure using Tree Tensor Networks, that exploits their hierarchical structure leading to a modest $O(\log N)$ computational scaling with system size. To showcase the applicability and efficiency of our method, we demonstrate the importance of magic in many-body systems via the following discoveries: (a) for one dimensional systems, we show that long-range magic displays strong signatures of conformal quantum criticality (Ising, Potts, and Gaussian), overcoming the limitations of full state magic; (b) in two-dimensional $\mathbb{Z}_2$ lattice gauge theories, we provide conclusive evidence that magic is able to identify the confinement-deconfinement transition, and displays critical scaling behavior even at relatively modest volumes. Finally, we discuss an experimental implementation of the method, which only relies on measurements of Pauli observables. 
\end{abstract}

\maketitle

\section{Introduction}

Over the last two decades, quantum information concepts have revolutionized the way we understand and approach the many-body problem \cite{nielsen2002quantum}. Remarkable insights on quantum matter have been obtained under the lens of entanglement, a measure of separability that has found applications over a wide range of phenomena, from real time dynamics \cite{Calabrese2005}, to topological order \cite{kitaev2006,levin2006} and classification of states \cite{chen2011,schuh2011}. 
A pivotal role in establishing these applications has been played by the development of trustful entanglement measures~\cite{RevModPhys.81.865}, in combination with efficient theoretical methods to explore that in the context of many-body systems~\cite{amico2008} - one paradigmatic example being tensor networks \cite{SCHOLLWOCK201196,ORUS2014117}.

On par with entanglement, another quantum information concept that is receiving increasing attention is that of non-stabilizerness, also known as magic \cite{bravyi2005UniversalQuantumComputation,campbell2017roads,bravyi2012magic}. In the context of quantum computing, magic is now understood as a fundamental resource that would be required to outperform classical simulations, and its concrete role in digital simulations has been widely addressed \cite{veitch2014resource,Wootters1987,Gross2006,Veitch2012,wang2019quantifying,wang2020efficiently}. However, differently from entanglement, there is presently limited understanding of how magic reflects many-body phenomena, and even if it does it at all~\cite{liu2022}: a fundamental limiting factor is that, oppositely to entanglement, we lack an array of scalable, efficient methods to actually compute magic - a shortage that severely limits our capability of identifying situations where there can be a direct connection between magic and physical phenomena.

In this work, we present a theoretical framework to measure many-body magic that leverages on a stochastic sampling of the system wave function. Our work builds upon recent developments in the field, in particular, on the recognition of stabilizer Renyi entropies (SREs) as measures of magic (including an experimental demonstration with 4 qubits) \cite{leone2022stabilizer,Oliviero_2022,leone2023,odavić2022complexity}. While a direct measure of the former is extremely challenging as it requires a number of measurements that grows exponentially with the size of the partition, we introduce a Markov chain on Pauli strings as a tool to distill the most relevant contribution to magic. We show that our protocol returns an unbiased estimator of SREs of all orders, and that it is efficient under several important scenarios: those include both full state magic (that is relevant, e.g., to quantify the overall difference from a stabilizer state), and long-range magic - a quantity that is akin to mutual information and that, crucially, is not plagued by any UV-divergences when applied to field theory.

The estimation of magic via Pauli-Markov chain is a general  construction, that is broadly applicable to computations as well as experiments. We explore in detail its capabilities in the context of tree tensor networks (TTN) ~\cite{gerster2014,Silvi2019}. At first, we perform extensive methodological checks, in particular, on the efficiency of Markov sampling and autocorrelations. We then showcase the flexibility of our approach with several applications, to understand advantages and overall comparison with recently introduced direct sampling methods that constitute the state of the art in terms of measuring many-body magic in numerical computations~\cite{haug2023quantifying,haug2023stabilizer,lami2023quantum}. 

Firstly, we consider one-dimensional systems. There, by considering both Ising, Potts and Heisenberg models, we show that full-state magic is not always indicative of quantum critical behavior. In particular, while it works for the conceptually simple cases of Ising (as already observed in Ref. \cite{oliviero2022ising,haug2023quantifying,haug2023stabilizer,lami2023quantum}) and Potts models, it spectacularly fails detecting any criticality in the case of spin-1 XXZ models. Oppositely, long-range magic (whose computation was not accessible before our algorithm, to the best of our knowledge) displays sharp signatures of critical behavior in all models considered. Our work thus clarifies how, in the context of critical behavior, it is fundamental to construct - and to compute - UV-divergence free estimators to understand the role of magic. 

Secondly, we consider two-dimensional interacting systems, where the connection between magic and many-body phenomena is uncharted territory. We focus on the $\mathbb{Z}_2$ lattice gauge theory, for two reasons: its importance as a paradigmatic model for more complicated lattice field theories, as it displays a confinement-deconfinement transition, as well as topological order; and its direct connection to the toric code, an epitome example of quantum memory based on the stabilizer language \cite{gottesman1997stabilizer,vedral1997quantifying,gottesman1998heisenberg,gottesman1998theory,aaronson2004improved,smith2006typical,horodecki2009quantum,gutschow2010entanglement,preskill2012quantum,harrow2017quantum}. Thanks to the very modest $O(\ln N)$ size-scaling of our algorithm versus system size $N$, we are able to consider systems up to 100 spins.
Our results show how both confined and deconfined phase have volume-law magic: most remarkably, magic features striking signatures of critical behavior. Close to the transition point, its behavior is akin to that of a Binder cumulant, as magic density displays a crossing as a function of volume, whose functional form is dictated by finite-size scaling theory. Even more remarkably, universal collapses are not only evident at modest volumes, but even at relatively small bond dimensions, signalling that magic might be considerably less affected than other observables by tensor network truncations. At the physical level, our results point out that magic may serve as an order parameter for confinement-deconfinement transitions, even at volumes where other quantities (e.g., order parameters) are of very limited use. 

Finally, we give a glimpse of the applicability of our approach to experiments. In that context, we discuss in detail experimental errors as a function of finite sampling, size, and autocorrelations. Our results indicate that the sampling needed to scale to large systems requires very fast repetition rates, which are available in solid state settings, but constitute a challenge for atomic experiments. 

The rest of the paper is structured as follows. In Sec.~\ref{sec:magic}, we review the basic properties of magic, as well as SREs. In Sec.~\ref{sec:monte_carlo}, we describe how to sample Pauli strings via Markov chains, discuss the efficiency of various estimators, and detail our implementation with tree tensor networks. In Sec.~\ref{sec:results}, we present our results on both one- and two-dimensional spin systems. In Sec.~\ref{sec:exp_protocol}, we detail our experimental protocol, and then conclude in Sec.~\ref{sec:concl}.

\section{Non-stabilizerness: Challenges in many-body physics}\label{sec:magic}

\subsection{Quick overview of resource theory for magic}

Quantum resource theories aim to capture the fundamental aspects inherent in quantum technology. For instance, entanglement is a crucial resource for quantum cryptography and communication. The resource framework for entanglement finds practical application by providing bounds on the efficiency of entanglement distillation protocols. Error-correcting codes play a fundamental role in achieving fault-tolerant quantum computation. These codes enable the storage of quantum information while protecting it from the detrimental effects of noise.

The development of error correcting codes based on the stabilizer formalism - e.g., the toric code - has motivated a resource theory of non-stabilizerness, or magic.
Here, we briefly review it, and summarize the main challenges in addressing magic in the context of many-body theory.

We first formally define the notation. We consider a system of $N$ qubits (generalizations to larger Hilbert spaces will be discussed below), with Hilbert space $\mathcal{H}=\otimes^N_{j=1} \mathcal{H}_j$. The $N$-qubit Pauli group $\mathcal{P}_N$ encompasses all Pauli string operators with an overall phase of $\pm i$ or $\pm 1$. Mathematically, we define $\mathcal{P}_N$ as follows:
\be 
\mathcal{P}_N = \left\lbrace e^{\frac{i \theta \pi}{2}} \sigma_{j_1} \otimes \dots \otimes \sigma_{j_N} | \theta, j_k = 0,1,2,3 \right\rbrace \, \, .
\ee
Moving on to stabilizer states, we can establish that a pure $N$-qubit state falls into this category if it satisfies certain conditions. Specifically, a stabilizer state is associated with an abelian subgroup $\mathcal{S}\subset \mathcal{P}_N$ that contains $2^N$ elements. For every $S\in \mathcal{S}$, the stabilizer state $\ket{\psi}$ remains unchanged under the action of $S$, expressed as $S|\psi\rangle=|\psi\rangle$. Alternatively, we can define stabilizers using Clifford unitaries, which are unitary transformations preserving the Pauli group when conjugated with it, i.e.
\be 
\mathcal{C}_N = \left\lbrace U \text{ s.t. } U P U^{\dagger} \in \mathcal{P}_N \, \text{for  all} \quad P \in \mathcal{P}_N  \right\rbrace \, .
\ee  
The Clifford set $\mathcal{C}_N$ can be generated using the Hadamard gate, the $\pi/4$-phase gate, and the CNOT gate. Notably, stabilizer states are pure quantum states that can be prepared by applying Clifford operations to a canonical trivial state $|0\rangle^{\otimes N}$.

In the framework of resource theory, stabilizer states are considered free states while Clifford unitaries and Pauli measurements
constitute free operations.  
The computation using only free states and free operations can be efficiently classically simulated, whereas universal quantum computation can be achieved through supplying magic (non-free) states.
Therefore to enable successful quantum computations, additional techniques are necessary to ensure the fault-tolerant implementation of a universal set of quantum gates. This can be achieved by augmenting the Clifford group with the Toffoli gate or the $\pi/8$ phase gate, thus unlocking the potential for universal quantum computation.

In this context, a central task is the quantification of the amount of non-Clifford operations needed to prepare a given quantum state. The properties required to a good measure $\mathcal{M}$ of non-stabilizerness are
(i) $\mathcal{M}(|\psi\rangle) =0$ $\Longleftrightarrow$ $| \psi \rangle$ is a stabilizer, (ii) non-increasing under Clifford operations: $\mathcal{M}(\Gamma |\psi\rangle) \leq \mathcal{M}(|\psi\rangle)$ if $\Gamma \in \mathcal{C}_N$, and (iii) $\mathcal{M}(|\psi\rangle \otimes |\phi\rangle) =  \mathcal{M}(|\psi\rangle)+ \mathcal{M}(|\phi\rangle)$.

For many-body systems, previous investigations into magic measures have primarily concentrated on small or weakly correlated systems, leading to a limited understanding of magic in entangled many-body systems.
An inherent challenge arises due to the exponential growth of stabilizer states and their increasingly intricate geometric structures as the system size expands. Consequently, the general calculation or numerical analysis of magic measures for large states becomes arduous. In order to enhance our understanding of this phenomenon, several fundamental questions require attention. These include comprehending the extent to which many-body quantum states can exhibit magic, determining the typical amount of magic found in generic states, and developing methodologies for computing the magic associated with many-body states.

From a quantum information viewpoint, the main motivation in understanding and measuring many-body magic stems from its relevance as a resource towards quantum advantage. Recent studies have shed light on the fact that the computational power of a state cannot be solely attributed to its magic density; other characteristics of magic may also play significant roles \cite{liu2022}. These properties encompass not only the primary aspect of magic density but also the subleading terms, nonlocal components, topological aspects, and more. Consequently, it is crucial to develop a numerical scheme capable of accessing and analyzing the various features of magic in many-body systems. Such a scheme would facilitate a comprehensive exploration and understanding of the intricate interplay between magic and computational power. This would constitute a major step forward in our endeavor to fully characterize the role played by magic in many-body systems. 

Notwithstanding such practical importance, understanding the connection between quantum correlations and physical phenomena is interesting from a broader perspective~\cite{liu2022} - especially, given the importance and impact such a connection has had in the case of entanglement. The connection between magic and physical phenomena is presently poorly understood, due to the combined lack of computable measures of magic, and of methods to attack them. 

From the point of view of observables, the key result we will exploit is Ref.~\cite{leone2022stabilizer}, that demonstrated SREs as a measure of magic (at least in the case of coherent dynamics; in more complicated scenarios, such quantities are not necessarily measures, see Ref.~\cite{haug2023stabilizer}). From the point of view of connection between magic and physical phenomena, three works are serving as a key motivation in this direction~\cite{haug2023quantifying,lami2023quantum,haug2023stabilizer}.
Thanks to the development of novel techniques based on direct sampling of matrix-product states (MPSs), these works have pointed out strong connections between critical behavior and magic in the context of one-dimensional systems, at precision and volumes never attained before. We will discuss this in more detail over the next section.

\subsection{Stabilizer Renyi entropy} \label{sec:sre}

Stabilizer Rényi Entropies (SREs) are a measure of nonstabilizerness recently introduced in Ref.~\cite{leone2022stabilizer}. For a pure quantum state $\rho$, SREs are expressed in terms of the expectation values of all Pauli strings in $\mathcal{P}_N$: 

\be \label{eq:SRE_def}
M_n \left( \rho\right)= \frac{1}{1-n} \log \left \lbrace \sum_{P \in \mathcal{P}_N} \frac{|\Tr \left(\rho P\right)|^{2n}}{d^N} \right \rbrace \  , 
\ee
with $d$ is the local dimension of the Hilbert space of $N$ qudits and $\mathcal{P}_N$ is the generalized Pauli group of $N$ qudits \cite{Gheorghiu2014}. The SREs have the following properties:   \cite{leone2022stabilizer} (i) faithfulness: $M_n(\rho)=0$ iff $\rho \in \text{STAB}$, (ii) stability under Clifford unitaries $C \in \mathcal{C}_N$: $M_n(C\rho C^\dagger)=M_n(\rho)$ , and (iii) additivity: $M_n(\rho_{A} \otimes\rho_{B})=M_n(\rho_{A})+M_n(\rho_{B} )$. The SREs are thus a good magic measure in the point of view of resource theory, where the free states are defined as the stabilizer states while the free operations are the Clifford unitaries. This definition is a straightforward generalization to general local dimension $d$ from the one given in Ref. \cite{leone2022stabilizer}. For $d>2$, the Pauli operators are no longer Hermitian, and thus the expectation values can be complex. In Eq. \eqref{eq:SRE_def}, we take the absolute values of the expectation values $|\Tr \left(\rho P\right)|$.  Eq. \eqref{eq:SRE_def} can be seen as the Rényi-$n$ entropy of the classical probability distribution:
\begin{equation}
    \Xi_P=|\Tr \left(\rho P\right)|^2/ d^N.
    \label{eq:Xi}
\end{equation}
It has the following properties:   \cite{leone2022stabilizer} (i) faithfulness: $M_n(\rho)=0$ iff $\rho \in \text{STAB}$, (ii) stability under Clifford unitaries $C \in \mathcal{C}_N$: $M_n(C\rho C^\dagger)=M_n(\rho)$ , and (iii) additivity: $M_n(\rho_{A} \otimes\rho_{B})=M_n(\rho_{A})+M_n(\rho_{B} )$. The SREs are thus a good magic measure in the point of view of resource theory, where the free states are defined as the stabilizer states while the free operations are the Clifford unitaries.

Moreover, the definition of SREs can be extended to mixed states by properly normalizing $\Xi_P$. For example, for $n=2$, the mixed state SRE is given by \cite{leone2022stabilizer}
\begin{equation} \label{eq:sre_mixed}
    \Tilde{M_2} = -\log \left(   \frac{\sum_{P \in \mathcal{P}_N}  |\Tr \left(\rho P\right)|^{4}}{\sum_{P \in \mathcal{P}_N} |\Tr \left(\rho P\right)|^{2}} \right),
\end{equation}
which can be seen as the Rényi-$2$ entropy of 
\begin{equation}
    \Tilde{\Xi}_P=|\Tr \left(\rho P\right)|^2/ {\sum_{P \in \mathcal{P}_N} |\Tr \left(\rho P\right)|^{2}},
    \label{eq:Xi_Tilde}
\end{equation}
apart from some offset. Here, the free states are defined as the mixed states that can be obtained from pure stabilizer states by partial tracing \cite{leone2022stabilizer}.

Furthermore, the long-range magic can be quantified by 
\begin{equation} \label{eq:lr_magic}
    L(\rho_{AB}) = \Tilde{M_2}(\rho_{AB}) - \Tilde{M_2}(\rho_A) - \Tilde{M_2}(\rho_B) 
\end{equation}
where $A$ and $B$ are two separated subsystems (see Fig. \ref{fig:schematic} (a)- (b)). A similar quantity has been considered previously in the context of mana \cite{white2021,Fliss2021} and robustness of magic \cite{Sarkar2020,bao2022}. $L(\rho_{AB})$ measures how magic is contained in the correlation between the subsystems, and thus it quantifies the degree to which magic cannot be removed by finite-depth quantum circuits \cite{white2021}. Indeed, due to the additivity of SRE, $L(\rho_{AB})$ vanishes for a product state $\rho_A \otimes \rho_B$. On the other hand, a non-vanishing value of $L(\rho_{AB})$ effectively quantifies the extent of deviation from the additivity in the case of entangled subsystems. 

The long-range magic is directly reminiscent of mutual information, that has played a major role in characterizing the distribution of both classical information and quantum correlations in many-body systems \cite{nielsen2002quantum,matsuda1996mutual,PhysRevB.82.100409,PhysRevLett.106.135701,wilms2011mutual,wilms2012finite,PhysRevLett.100.070502,casini2007mutual,PhysRevResearch.4.033212,PhysRevLett.130.021603,calabrese2004entanglement,caraglio2008entanglement,furukawa2009mutual}. 
On the lattice, the main motivation for looking at functionals such as in Eq.~\eqref{eq:lr_magic} is that they are much more meaningful than simple bipartition properties from a field theory standpoint. Indeed, these quantities are expected to be free of UV divergences, and thus solely dominated by infrared, universal properties of the lattice theory. This parallels the f-functions used in field theory~\cite{casini2004finite}.

As discussed above, SREs have attracted recent interest due to their computability. The first technique was introduced in \cite{haug2023quantifying}, which expressed the SREs of integer index $n>1$ as the norm of a ``$2n$-replica'' MPS.  Although this technique yields an exact value of $M_n$ within a given MPS, its computational cost scales as a large power of the bond dimension $\chi$, specifically $O(N\chi^{6n})$. Thus, although the method is efficient in principle, in practice it can only access bond dimension up to $\chi=12$, which limits its applicability to investigate many-body physics. 

A different approach based on sampling of Pauli strings according to the probability distribution $\Xi_P$ was proposed very recently in \cite{lami2023quantum,haug2023stabilizer}. 
In those works, the Pauli strings are sampled directly via the perfect sampling scheme with Matrix Product State (MPS) introduced in \cite{ferris2012}. For the case of open boundary conditions (OBCs), the cost scales as $O(N\chi^3)$, thus enabling access to larger bond dimensions, which opens the door for investigating magic in entangled states. However, as we discuss in more detail in the next section, this method provides only an efficient estimation of $M_1$. It has been demonstrated that for $0 < n < 2$, $M_n$ violates monotonicity under measurements followed by conditioned Clifford transformations \cite{haug2023stabilizer}. Thus, it is important to develop an efficient scheme to efficiently compute $M_2$, which also has the nice property of being experimentally measurable \cite{leone2023,haug2023}. Furthermore, $M_2$ is directly linked to the average over the Clifford orbit of entanglement spectrum flatness in an arbitrary bipartition \cite{tirrito2023} and participation entropy flatness \cite{turkeshi2023measuring}. 

We also note that the aforementioned two methods have inherent limitations when it comes to evaluating magic within a subsystem of a state - for instance, none can access long-range magic. As a result, the existing techniques are unable to provide insights into how magic is distributed within a given state. 

\begin{figure}
    \centering
    \includegraphics[width=0.9\linewidth]{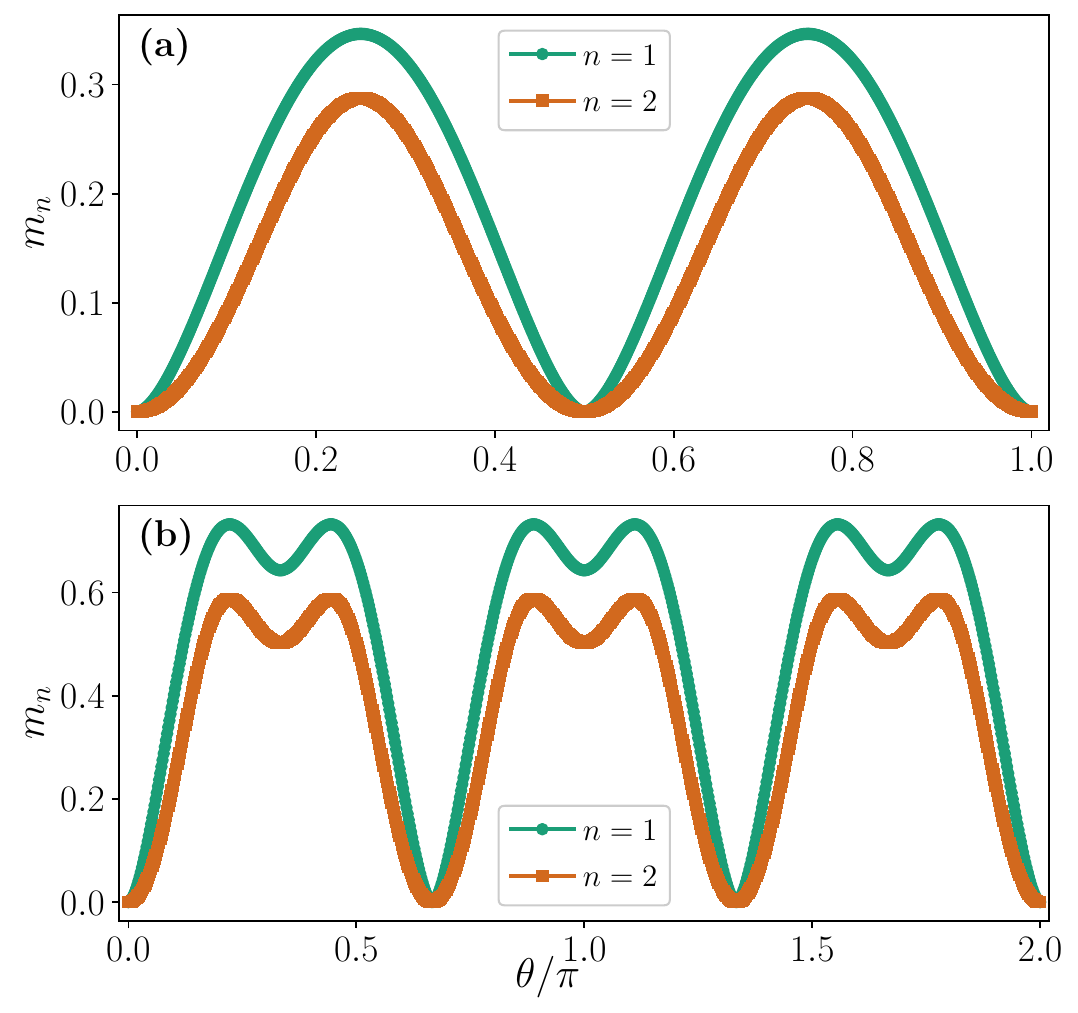}
    \caption{
    \textbf{Stabilizer entropies for qubit and qutrit.}
    The SRE density $m_1$ and $m_2$ for single qubit state (a) defined in the Eq. (\ref{eq:single_qubit_state}), and for single qutrit state defined in Eq. (\ref{eq:single_qutrit_state}) (b). }
    \label{fig:SREs_examples}
\end{figure}

\subsubsection{Examples}
To familiarize with the behavior of SREs in many-body systems, here we provide some examples of SREs in simple wave functions. First of all, we stress that the SREs are basis-dependent, i.e., it is not invariant under local basis change. In particular, the SREs of a single-qubit state may be non-trivial. For example, consider the following one-parameter family of single-qubit states
\begin{equation} \label{eq:single_qubit_state}
|\psi (\theta)\rangle= \frac{1}{\sqrt{2}} \left[ |0 \rangle + e^{i\theta} |1\rangle  \right]\, .
\end{equation}
Note that $|\psi (\pi/4)\rangle$ corresponds to the canonical T-state. The SREs can be computed easily by evaluating the expectation values of $P \in \{I,X,Y,Z \}$, and then plugging it in Eq. \eqref{eq:SRE_def}. The result is shown in Fig. \ref{fig:SREs_examples} (a). As can be seen, the SREs are non-zero apart from some special points $\theta=m \pi/2$ with integer $m$.

Now, the SREs of a product state of $N$ copies of $|\psi (\theta)\rangle$ can also be computed straightforwardly, utilizing the additivity property of SRE, $M_n(\rho_{A} \otimes\rho_{B})=M_n(\rho_{A})+M_n(\rho_{B} )$. The SREs are then just given by $M_n(|\psi (\theta)\rangle^{\otimes N})=N M_n(|\psi (\theta)\rangle)$.

For an example of qudit states, we consider the following family of single-qutrit states
\begin{equation} \label{eq:single_qutrit_state}
|\phi (\theta)\rangle= \frac{1}{\sqrt{3}} \left[ |0 \rangle + e^{i\theta}  |1\rangle + e^{-i\theta} |2\rangle  \right]\, .
\end{equation}
Here, $|\phi (2\pi/9)\rangle$ corresponds to the canonical qutrit T-state. We now need to compute the expectation values of $3^2$ single-qutrit Pauli operators. To define the Pauli operators, we first define the shift and clock operators for $d$-level system as
\be \label{eq:clock_shift}
X = \sum_{k=0}^{d-1} |k+1\rangle \langle k | \quad \textrm{and} \quad Z=\sum_{k=0}^{d-1} \omega^k_d |k\rangle \langle k |,
\ee
 where $\omega_d=e^{2\pi i/d}$, and the addition is defined modulo $d$. For qutrits, we have $d=3$. 
 The qudit Pauli operators  are defined as
\begin{equation} \label{eq:pauli_qudit}
    T_{aa'} = \omega^{-2^{-1}aa'}Z^a X^{a'}
\end{equation}
for $a,a' \in \mathbb{Z}_d$. Here, $2^{-1}$ is the inverse element of $2$ in $\mathbb{Z}_d$.

Computing the expectation values of the Pauli operators in Eq. \eqref{eq:pauli_qudit}, we can compute the SREs of $|\phi (\theta)\rangle$ using Eq. \eqref{eq:SRE_def}. The result is shown in Fig. \ref{fig:SREs_examples} (b). In this case, the SREs are non-trivial apart from some special points $\theta=m 2\pi/3$ with integer $m$.

\section{Markov chain Monte Carlo sampling of Pauli strings}
 \label{sec:monte_carlo}      
 In this work, we investigate the SREs using Monte Carlo sampling of Pauli strings according to some probability distribution $\Pi_P$, which only depends explicitly on the expectation values of Pauli strings. 
 For example, for the calculation of $M_n$, we get $\Pi_P = \Xi_P$ (Eq.~\eqref{eq:Xi}), while for $\tilde{M}_2$ we have $\Pi_P = \tilde{\Xi}_P$ (Eq.~\eqref{eq:Xi_Tilde}).
Here we focus on Metropolis algorithm, although other sampling methods, such as heat bath, may also be employed. Since $\Pi_P$ only depends on the expectation value of $P$, this method is applicable to any numerical methods in which expectation values of (non-local) operators can be accessed, such as exact diagonalization and tensor network methods. Furthermore, this method can also be utilized to experimentally measure SREs (see Sec. \ref{sec:exp_protocol}).

\begin{algorithm}[H] 
\caption{Monte Carlo sampling of Pauli strings} \label{alg:QA_TTN}
\begin{flushleft}
 \textbf{Input}: a quantum state $\rho$ and number of sampling $N_S$ 
\end{flushleft}
\begin{algorithmic}[1]
%\State Put the TTN in central canonical form.
\State Initialize the Pauli string $P$.
\State Compute $\Tr (\rho P)$ and $\Pi_P$.
\For{($i=1$; $i \leq N_S$; $i++$)}
     \State Propose a candidate Pauli string $P^{\prime}$.
     \State Compute $\Tr (\rho P^{\prime})$ and  $\Pi_{P'}$.
     \State  Accept the move with probability: $\min \left( 1,\frac{\Pi_{P'}}{\Pi_P} \right)$.
     \State Measure the estimators.
\EndFor
\end{algorithmic}
\begin{flushleft}
\textbf{Output}: a Markov chain of $P$ with probability $\Pi_P$.
\end{flushleft}
\end{algorithm}

\subsection{Algorithm theory}

The scheme is summarized in the Algorithm~\ref{alg:QA_TTN}.
If we sample according to $\Xi_P$, $M_n$ can be estimated using the unbiased estimators
\be \label{eq:estimator_n}
M_n = \frac{1}{1-n} \log  \left\langle |\Tr (\rho P)|^{2(n-1)} \right\rangle_{\Xi_P} 
\ee
for $n > 1$ and 
\be \label{eq:estimator_1}
M_1 =   \left\langle -\log \left( |\Tr(\rho P)|^{2} \right) \right\rangle_{\Xi_P}
\ee
for $n=1$, where $\langle ... \rangle_{\Xi_P}$ is the average over $\Xi_P$ obtained with sampling. For $n<1$, a better estimation can be done by reversing Eq. \eqref{eq:estimator_n}, i.e.,
\be \label{eq:estimator_n<1}
M_n = -\frac{1}{1-n} \log  \left\langle |\Tr (\rho P)|^{2(1-n)} \right\rangle_{\Pi_{P,n}} 
\ee
where $\Pi_{P,n} \propto |\Tr(\rho_{AB} P)|^{2n}$.
Let us analyze the efficiency of these estimators.

 \paragraph*{\ding{233} SRE with $n=1$.--} For $n=1$, the variance of $M_1$ is shown to be at most quadratic in $N$ in Ref. \cite{lami2023quantum}. Thus, the estimator for $M_1$ is efficient. Actually, we can even make a stronger statement, if we make the assumption that the SREs are linear in $N$, i.e., $M_\alpha=Nf(\alpha)+O(1)$, where $f(\alpha)$ is a function that does not depend on $N$. Using the relation~\cite{deboer2019}: 
 \begin{equation}
    \textrm{Var} (M_1)=\frac{d^2[(1-\alpha)M_\alpha]}{d\alpha^2} \Bigg|_{\alpha=1},
\end{equation}
we see that $\textrm{Var} (M_1)$ is linear in $N$. It follows that the variance (standard deviation) of the SRE density, $m_1=M_1/N$, scales as $1/N$ ($1/\sqrt{N}$).

\paragraph*{\ding{233} SRE with $n \neq 1$.--} For $n> 1$, the variance of Eq. \eqref{eq:estimator_n} is given by
\begin{equation} \label{eq:var}
\begin{split}
     \mathrm{Var} & \left(|\Tr(\rho P) |^{2(n-1)} \right)  \\ 
     &= \left\langle |\Tr (\rho P )| ^{4(n-1)}\right\rangle_{\Xi_P} - \left\langle |\Tr (\rho P) \rangle|^{2(n-1)}\right\rangle_{\Xi_P}^2 \\
     &= \exp \left[ -2(n-1)M_{2n-1} \right] - \exp \left[ -2(n-1)M_{n} \right]. 
\end{split} 
\end{equation}
Now, by second-order approximation $\textrm{Var} \left( \log x\right)\approx \mathrm{Var} \left( x\right)/x^2$, we have
\begin{equation}
\begin{split}
    \mathrm{Var} \left(M_n \right) &\approx \frac{\exp \left[ -2(n-1)M_{2n-1} \right] - \exp \left[ -2(n-1)M_{n} \right]}{|n-1|\exp \left[ -2(n-1)M_{n} \right]}  \\
     & = \frac{\exp \left[ 2(n-1)(M_n-M_{2n-1}) \right] - 1}{|n-1|}. 
\end{split}
\end{equation}
For $n<1$, 
\begin{equation} \label{eq:varn<1}
\begin{split}
    \textrm{Var} &\left(|\Tr(\rho P) |^{2(1-n)} \right) \\&= \left\langle |\Tr(\rho P) |^{4(1-n)}\right\rangle_{\Pi_{P,n}} - \left\langle |\Tr(\rho P) |^{2(1-n)}\right\rangle_{\Pi_{P,n}}^2 \\
     &= \exp \left[ (n-1)(M_{2-n}+M_n) \right] - \exp \left[ 2(n-1)M_{n} \right]. 
\end{split}
\end{equation}
Then,
\begin{equation}
\begin{split}
    \textrm{Var} \left(M_n \right) &\approx \frac{\exp \left[ (n-1)(M_{2-n}+M_n) \right] - \exp \left[ 2(n-1)M_{n} \right]}{|n-1|\exp \left[ 2(n-1)M_{n} \right]}  \\
     &= \frac{\exp \left[ (1-n)(M_n-M_{2-n}) \right] - 1}{|n-1|}. 
\end{split}
\end{equation}
In both cases, if the SREs grow at most logarithmically in $N$, the variance grows at most polynomially. Thus, by Chebyshev's inequality, the number of samples needed for a fixed error $\epsilon$ is polynomial in $N$, i.e, the estimator is efficient. On the other hand, if the SREs are linear in $N$, as is typically the case in many-body systems \cite{oliviero2022ising,white2021, haug2023quantifying}, the variance grows exponentially with $N$ when $n\neq 1$. Thus, the estimator for $M_n, n\neq 1$ is efficient only if the SREs are at most $O(\log N)$. One can also see this intuitively by noting that the quantity being estimated is exponentially small in $N$ when $M_n$ is linear, and thus we need exponentially small precision. We note in passing that states with logarithmically growing SREs can arise in many-body systems in the frustrated regime \cite{odavić2022complexity}.

Note, however, that the SREs are typically linear in $N$. Therefore, using the estimators in Eq. \eqref{eq:estimator_n}, the estimation of $M_n, n \neq 1$ will almost always be exponentially costly. Nevertheless, the cost typically grows much more slowly than $d^{2N}$ which is the cost for exact computation. Thus, in practice, using this estimator is still beneficial to extend the system sizes we can study, as we shall illustrate in Sec. \ref{sec:results}. Importantly, using Monte Carlo sampling, we are not restricted to sample the Pauli strings according to $\Xi_P$. An alternative approach is to sample Pauli strings according to the probability distribution $\Pi_{P,n} \propto \Tr(\rho P)^{2n}$. We then need to estimate the normalization constant of $\Pi_{P,n}$ to estimate $M_n$. This is a non-trivial task, equivalent to estimating the partition function, for which a wealth of sophisticated methods have been put forward \cite{Zwanzig1954,Bennett1976,Chib1995, Gelfand1994,Diciccio1997,Neal2001,Meng1996SIMULATINGRO,Meng1996,Gelman1998,Chen2000,wang2001,troyer2003}.

\begin{figure}
    \centering
    \includegraphics[width=0.9\linewidth]{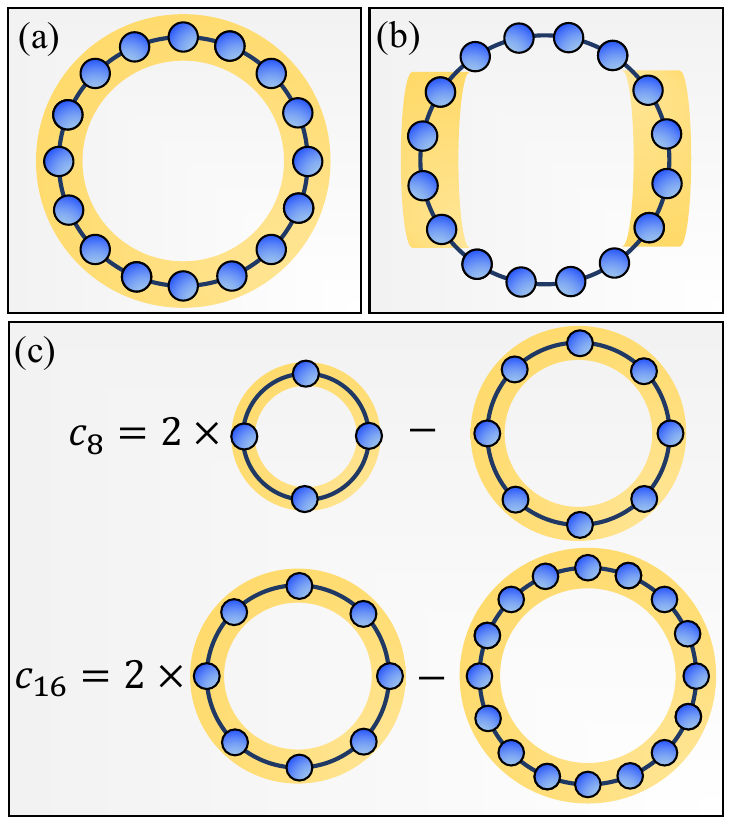}
    \caption{
    \textbf{Schematics of partitions.}
    (a) Full partition. (b) Two widely-separated partitions for the calculation of long-range magic in Eq. \eqref{eq:lr_magic}. (c) Subleading term as in Eq. \eqref{eq:subleading}, as well as a cartoon depicting the increment trick discussed in the main text.}
    \label{fig:schematic}
\end{figure}

\paragraph*{\ding{233} Long-range magic.--} In addition, we are interested to estimate the long-range magic as quantified by $L(\rho_{AB})$ in Eq. \eqref{eq:lr_magic}. While we can in principle compute the individual $\Tilde{M_2}$ for $\rho_C, C\in \{A,B,AB\}$, this is not optimal, as we have seen that the estimation for $\Tilde{M_2}$ is not efficient when $\Tilde{M_2}$ grows linearly with $N$. Moreover, we expect that the leading term of $\Tilde{M_2}$ will be canceled out in $L(\rho_{AB})$. In this case, it is more desirable to estimate $L(\rho_{AB})$ directly, without having to resort to inefficient estimation of $\Tilde{M_2}$. To do this, we first rewrite Eq. \eqref{eq:lr_magic} as follows:
\begin{equation}
    L(\rho_{AB}) = I_2(\rho_{AB}) - W(\rho_{AB}),
\end{equation}
where 
\begin{equation}
\resizebox{.89\hsize}{!}{$W(\rho_{AB}) = -\log \left( \frac{\sum_{P_A \in \mathcal{P}_A} |\Tr(\rho_A P_A)|^4 \sum_{P \in \mathcal{P}_B} |\Tr(\rho_B P_B)|^4}{\sum_{P_{AB} \in \mathcal{P}_{AB}}  |\Tr(\rho_{AB} P_{AB})|^4} \right)$},
\end{equation}
and $I_2(\rho_{AB})=S_2(\rho_{A}) + S_2(\rho_{B}) - S_2(\rho_{AB})$ is the Rényi-2 mutual information. If one is to sample according to $\Pi_{P_{AB}} \propto \Tr(\rho_{AB} P_{AB})^4$, we can estimate $W(\rho_{AB})$ by
\begin{equation} \label{eq:estimator_w}
    W(\rho_{AB}) =  -\log  \left\langle \frac{ |\Tr  (\rho_A P_A)|^4 |\Tr  (\rho_B P_B)|^4 }{|\Tr  (\rho_{AB} P_{AB})|^4}\right\rangle_{\Pi_{P_{AB}}}, 
\end{equation}
where $P_{AB}$ is decomposed as $P_{AB}=P_A \otimes P_B$. Similarly, we have
\begin{equation} \label{eq:estimator_i2}
I_2(\rho_{AB}) =  -\log  \left\langle \frac{| \Tr  (\rho_A P_A)|^2 |\Tr  (\rho_B P_B)|^2 }{|\Tr  (\rho_{AB} P_{AB})|^2}\right\rangle_{\Xi_{P_{AB}}}.
\end{equation}
Therefore, as a byproduct, our scheme can be applied to compute the Rényi mutual information for disjoint subsystems.

\paragraph*{\ding{233} Subleading term.--} The previous scheme can be straightforwardly modified to extract the subleading term in the expansion $M_n(N)=D_N N+c_N$ \cite{haug2023quantifying}. Here we consider 1D systems for simplicity. Specifically, the subleading term is approximated by the quantity $c_N=2M_n(N/2)-M_n(N)$ (see Fig. \ref{fig:schematic} (c)), which expands as
\begin{equation} \label{eq:subleading}
    c_N =  \log  \left\langle \frac{ |\Tr  (\rho_{N/2} P^{(1)})|^{2n} |\Tr  (\rho_{N/2} P^{(2)})|^{2n} }{|\Tr  (\rho_{N} P)|^{2n}}\right\rangle_{\Pi_{P,n}} 
\end{equation}
for $n \neq 1$, where $\rho_{N,N/2}$ is the density matrix for a 1D system of size $N$ and $N/2$, respectively. 
For simplicity, we have assumed translational invariance in Eq.~\eqref{eq:subleading}, but the procedure can be straightforwardly generalized to any system.
Here, denoting $P=P_1 P_2 ... P_N$, where $P_i$ is a Pauli operator acting on site $i$ in the $N-$site system, we choose $P^{(1)}=P_1 P_2 ... P_{N/2}$ and $P^{(2)}=P_{N/2+1} P_{N/2+2} ... P_{N}$. Note that, differently from Eq. \eqref{eq:estimator_w}, here we consider two pure states of different sizes $N$ and $N/2$. 
For the subleading term in 1D systems, the term inside the log in Eq.~\eqref{eq:subleading} does not decay exponentially, and thus the estimation can be done more efficiently than the estimation of the leading term in Eq.~\eqref{eq:estimator_n}. 

\paragraph*{\ding{233} Increment trick for SRE.--}
The extraction of the subleading term in Eq. \eqref{eq:subleading} presents an alternative strategy to estimate $M_n$, which circumvents the problem of exponential variance for the estimator in Eq. \eqref{eq:estimator_n}. The key idea is that, if the estimation in Eq. \eqref{eq:subleading} is efficient, then we can estimate $c_N,c_{N/2}, ...,$ until the size is small enough that $M_n$ can be evaluated exactly. The number of $c_M$'s that needs to be computed scales as $O(\log N)$ (assuming translational invariance). Then, we can determine $M_n(N)$ by considering a proper linear combination of $c_M$'s. This strategy is reminiscent of the increment trick employed in estimation of Rényi entanglement entropies in Quantum Monte Carlo simulations \cite{hastings2010,humeniuk2012,Zhao2022}, which considers the difference of Rényi entropies of smaller and smaller regions, to compute the Rényi entropy of a large entangling region with high precision. However, in this case, the form of $c_N$ is specifically designed to cancel out the volume-law term of $M_n$, differently from entanglement entropy which exhibits area law. 

The above strategy is effective in 1D systems because the subleading term $c_N$ is expected to either remain independent of system size or exhibit at most logarithmic growth. However, in higher-dimensional systems, $c_N$ may exhibit area-law scaling, leading to growth with size. In this case, more complicated linear combination of $M_n$'s shall be considered to eliminate the area-law term (while, at the same time, also keeping the volume law one vanishing). For example, in 2D systems, the form of linear combination used in extracting the topological entanglement entropy with Kitaev-Preskill \cite{kitaev2006} or Levin-Wen scheme \cite{levin2006} will cancel both the volume-law and area-law term. It is convenient to partition the system into four subsystems as proposed in \cite{gerster2017}, which is also suitable with 2D TTN geometry. With this scheme, the estimation of $M_n(L \times L)$ is reduced to $M_n(L/2 \times L), M_n(L/2 \times L/2),...$, such that only $O(\log N)$ computations are required, as in 1D case\footnote{In each computation, one estimates the linear combination $\gamma_{M_n}=M_n(L \times L) - 4M_n(L/2 \times L)+4M_n(L/2 \times L/2)$, and similar form for $L/2 \times L$ geometry. This quantity can be recast into a form suitable for Monte Carlo estimation in a similar way as the subleading term in Eq. \eqref{eq:subleading}.}.

\begin{figure*}[htb] 
    \centering
    \includegraphics[width=1\linewidth]{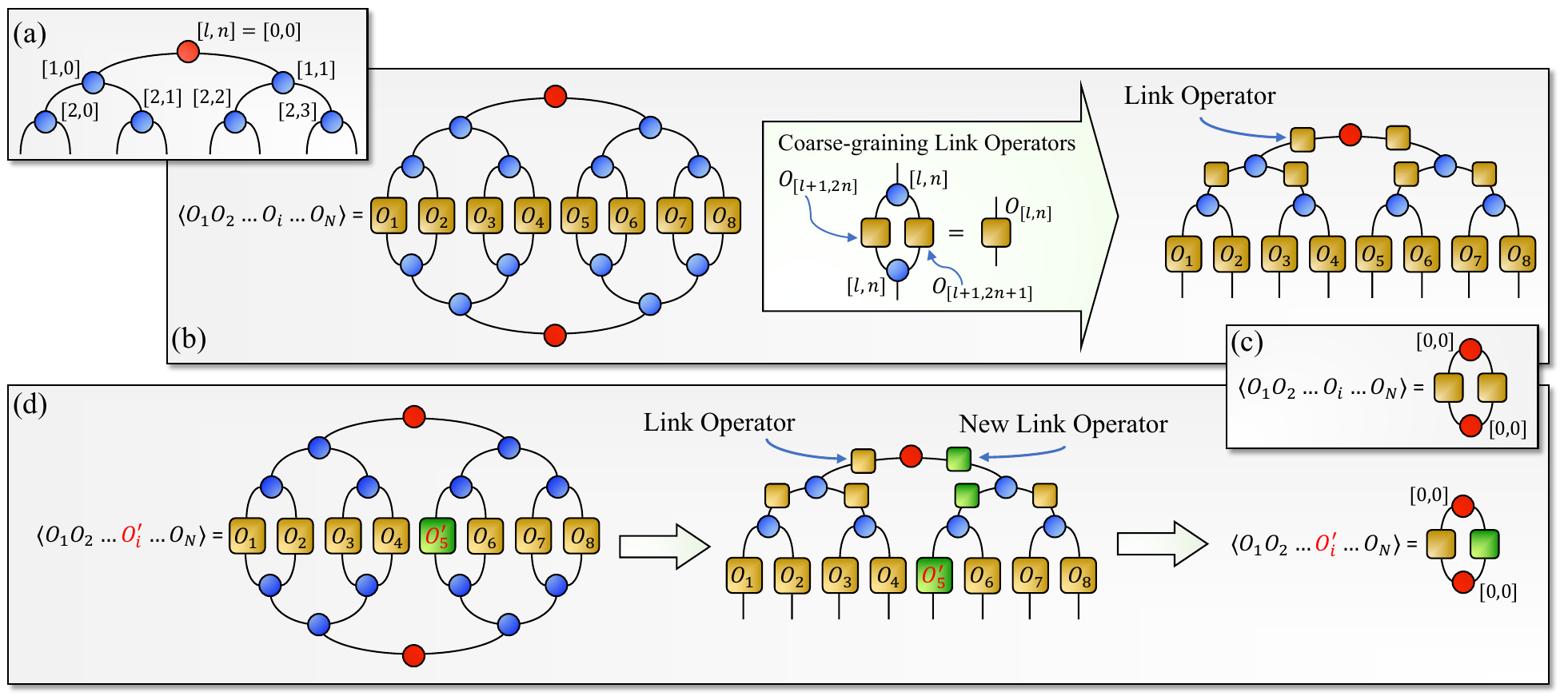}
    \caption{
    \textbf{Efficient Monte Carlo sampling using Tree Tensor Network.}
    (a) Tree Tensor Network (TTN) representation of a many-body wavefunction $| \psi \rangle$, where tensors are depicted as circles arranged in a binary-tree structure. Each tensor is identified by a pair of zero-indexed integers $[l, n]$, representing its layer index $l$ and tensor index $n$ at that layer. The red circle at the top-most layer represents the root tensor having index $[0, 0]$,  where the isometry center of the TTN is taken. 
    (b) To evaluate the expectation value of a tensor-product of single-site operators $\langle O_1 O_2 \ldots O_N \rangle$, we first place each operator $O_i$ at the physical site it acts on in the TTN representation. Then, we compute the effective link operators which live at the virtual links by the coarse-graining procedure as shown in the figure. The coarse-graining is performed iteratively from the physical sites to the top-most virtual links, which are directly connected to the root tensor. 
    At each step, the link operators $O_{[l+1, 2n]}$ and $O_{[l+1, 2n+1]}$ are combined into $O_{[l, n]}$ by the $[l, n]$-tensor. The resulting link operator $O_{[l, n]}$ acts on the $[l-1,\lfloor n/2 \rfloor]$-tensor one layer above in the TTN structure.
    (c) The expectation value  $\langle O_1 O_2 \ldots O_N \rangle$ is calculated from the contraction of the root $[0, 0]$-tensor and the top-lost link operators as shown in the figure. 
    (d) Considering a modified operator which differs only at a single site from the previous one, $ O_1 O_2 ... O'_i ... O_N$, we only need to recompute the link operators in the path from the modified physical site $i$ to the topmost link.}
    \label{fig:ttn}
\end{figure*}

\subsection{Efficient sampling with tensor networks: the example of tree tensor networks}
 \label{sec:ttn_sampling}

The probability $\Pi_P$ of a given Pauli string $P$ only depends on the expectation value of $P$, and thus it is efficiently computable in TTN (or any loopless tensor network \cite{Silvi2019}).
Following the convention introduced in Ref.~\cite{gerster2014}, each tensor in the TTN structure is denoted by the pair of zero-indexed integers $[l, n]$, where $l$ corresponds to the layer index (starting from the top root tensor) and $n$ denotes the tensor at a particular layer $l$ counted from the left (see Fig.~\ref{fig:ttn}(a)). Obviously, in this notation, the top root tensor is represented by $[l, n]=[0, 0]$.

The algorithm to sample Pauli strings for the ground state of a quantum many-body system is described below.
\begin{itemize}
    \item[\ding{233}] After performing the adaptive variational ground-state search~\cite{gerster2014} for a many-body Hamiltonian, we arrive at the TTN representation of the many-body ground state wavefunction $\ket{\psi}$. We start by bringing the TTN into the central canonical form, where the $[0, 0]$-tensor is the orthogonality center (see Fig.~\ref{fig:ttn}(a)).
    \item[\ding{233}] Given the initial Pauli string $P=P_1 P_2 \ldots P_N$, where $P_i$ is a Pauli operator at site $i$, we construct the coarse-grained effective ``link'' operators $O_{[l, n]}$ at each link iteratively from the physical sites to the top-most links, where at the bottom-most (i.e., the physical) layer these link operators are identified with the Pauli operators (see Fig.~\ref{fig:ttn}(b)). At each step,  the link operators $O_{[l+1, 2n]}$ and $O_{[l+1, 2n+1]}$ are coarse-grained into $O_{[l, n]}$ by $[l, n]$-tensor. The new link operator $O_{[l, n]}$ acts on the $[l-1,\lfloor n/2 \rfloor]$-tensor a layer above in the TTN structure. We keep all the link operators in memory for future uses.
    \item[\ding{233}] The expectation value $\langle \psi | P | \psi \rangle$ now only involves the root $[0,0]$-tensor and top-most link operators $O_{[1, 0]}$ and $O_{[1, 1]}$ as seen in Fig.~\ref{fig:ttn}(c). 
    \item[\ding{233}] At each sampling step, we either propose a single-site update $P' = P_1 \ldots P_i' \ldots P_N$, or a two-site update $P' = P_1 \ldots P_i' \ldots P_j' \ldots P_N$, following Algorithm~\ref{alg:QA_TTN}. The updated sites $i$ and  $j$ are chosen randomly.
    \item[\ding{233}] We observe (Fig.~\ref{fig:ttn}(d)) that the effective link operators for $P'$ only differ with those of $P$ on the links that lie on the path from the site $i$ (or $j$) to the root $[0, 0]$-tensor. The number of such links scales only logarithmically in system size. This implies that computing $\langle \psi | P' | \psi \rangle$ can be done very efficiently with a computational cost of $O(\log(N)\chi^4)$, as opposed to $O(N\chi^4)$ for a generic many-body operator for the TTN. 
\end{itemize}

The heart of our efficient sampling procedure lies within the above observation for TTN. We exploit this scaling property to perform efficient Monte Carlo sampling of Pauli strings by standard Metropolis algorithm, where the candidate Pauli string for the next configuration only differs at a few sites with the previous Pauli string configuration. Crucially, the sites can be chosen arbitrarily, and this does not change the $\log N$ scaling of the TTN sampling, provided that the number of modified sites does not scale with system size. This allows for flexible sampling strategy, which can be designed by taking into account our knowledge about the state that we want to sample  -- very much like Monte Carlo methods are designed to probe partition functions.

The final step for calculating the expectation value of a proposed candidate Pauli string at each Metropolis iteration is the following.
\begin{itemize}
    \item[\ding{233}] 
    The link operators, that reside in the path from the updated site $i$ (or $j$) to the root $[0, 0]$-tensor, are updated by the coarse-graining step. The expectation value $\langle \psi | P' | \psi \rangle$ is now calculated by tensor contractions of the root tensor and top-most (updated) link operators (see Fig.~\ref{fig:ttn}(d)).
\end{itemize}

.

At this stage, it is important to discuss the efficiency of the more widely used MPS tensor network structure in relation to our sampling strategy. The  computational cost for direct sampling of Pauli strings using MPS with OBCs scales as $O(N \chi^3)$~\cite{lami2023quantum, haug2023stabilizer, ferris2012}, that also holds for Monte Carlo sampling using MPS\footnote{Using the MPS structure, the cost of each iteration in Monte Carlo sampling using single-site update can be reduced from $O(N)$ to $O(1)$ in $N$ using sequential left $\leftrightarrow$ right sweeps of updates, but this strategy trivially induces exploding autocorrelation time, making it unusable for practical purposes. Moreover, this sequential strategy becomes a real problem for two-site updates that are required for systems that preserves some symmetries (see Sec.~\ref{sec:results}).}, as opposed to the $O(\log(N) \chi^4)$ that we get utilizing TTN. 
Consequently, our method with TTN for obtaining SREs becomes increasingly efficient as the number of qudits $N$ grows large, particularly when $N/\log N \gtrsim \chi$.
Specifically, since the MPS or the TTN bond dimension $\chi$ saturates to a constant value with $N$ in 1D quantum systems with gapped spectrum  due to the area-law of entanglement entropy, our approach involving TTN vastly outperforms MPS based methods in terms of efficiency for large $N$.
Most importantly, the enhanced connectedness inherent in the TTN structure allows for efficient exploration of higher-dimensional (2D and even 3D) many-body systems (see e.g.,~\cite{tagliacozzo2009, Cataldi2021, Felser2020, Magnifico2021, Felser2021}). This paves the way to investigate SREs in higher-dimensional systems, as we present in Sec. \ref{sec:results}.

Finally, we mention that our scheme can also be used to compute the SREs of any partition of the system. To do this, we only need to restrict the Pauli strings to have support on the sites in the partition. Using the estimator for $n=2$, the same Monte Carlo procedure will yield $\Tilde{M_2}$ in Eq. \eqref{eq:sre_mixed}. Moreover, the algorithm is easily generalized to Tree Tensor Operator (TTO) \cite{Arceci2022}, which represents many-body density operator for mixed states.  

We note that the use of Monte Carlo techniques in tensor network has been considered before \cite{Schuh2008,Sandvik2007,ferris2012_2} to compute the expectation value of a local operator. Instead, here the expectation values are computed exactly, while the sampling is done at the level of operators being computed.

\section{Application to Quantum Many-Body Systems} \label{sec:results}

We apply the TTN based sampling method  in Sec.~\ref{sec:ttn_sampling} using the estimators in Eq. \eqref{eq:estimator_n} and Eq. \eqref{eq:estimator_1} to investigate the SREs in various many-body systems, especially near quantum critical points, both in 1D and 2D geometries. 
Unlike MPS, the structure of TTN allows for efficient exploration of systems under periodic boundary conditions (PBC) with similar computational cost as the open boundary conditions~\cite{gerster2014}. Therefore, we consider the periodic many-body systems, i.e., ring and torus geometry in 1D and 2D, respectively, to avoid boundary effects.
For the analysis of statistical errors and the autocorrelation times in the Markov chain samples, we refer to the Appendix~\ref{sec:auto_corr}, whereas for the analysis of convergence with bond dimension of the TTN, we refer to the Appendix ~\ref{sec:bond_dim}.

To obtain the TTN representation of the ground state of many-body systems we perform variational minimization with TTN sweeping algorithm \cite{gerster2014,Silvi2019}, and then employ the sampling scheme in Sec.~\ref{sec:ttn_sampling} to estimate the SREs of the ground state.  In particular, since the SREs are generally linear in the number of qudits $N$, we focus on the SRE densities $m_n=M_n/N$. 

All of the models we consider possess $\mathbb{Z}_n$ symmetry, with $n = 2$ or $3$, and thus, a two-site update scheme is required to sample only the Pauli strings that preserve the symmetry. 
The Pauli strings that preserve the $\Z_n$ symmetry, generated by $\prod_i Z_i$, are generated by $Z_i$ and $X_i^\dagger X_j$ (up to a phase constant). Here, $X$ and $Z$ are the shift and clock operators defined in Eq. \eqref{eq:clock_shift}  To ensure that only the Pauli strings that obey the $\Z_n$ symmetry are  considered, we generate the candidate Pauli string  $P'$ by randomly multiplying the current Pauli string $P$ with either $Z_i$ or $X_i^\dagger X_j$. It is easy to see that the update scheme is ergodic. For $d=3$, we set the probability to multiply with $Z_i$ or $Z_i^\dagger$ to be equal, so as to satisfy detailed balance. For $d=2$, when there is time-reversal-symmetry, the Pauli strings are additionally constrained to those with even numbers of $Y=iZX$. As such, the Pauli strings with odd numbers of $Y$ can be directly rejected.

\subsection{Non-stabilizerness in 1D many-body systems}

\begin{figure}
    \centering
    \includegraphics[width=\linewidth]{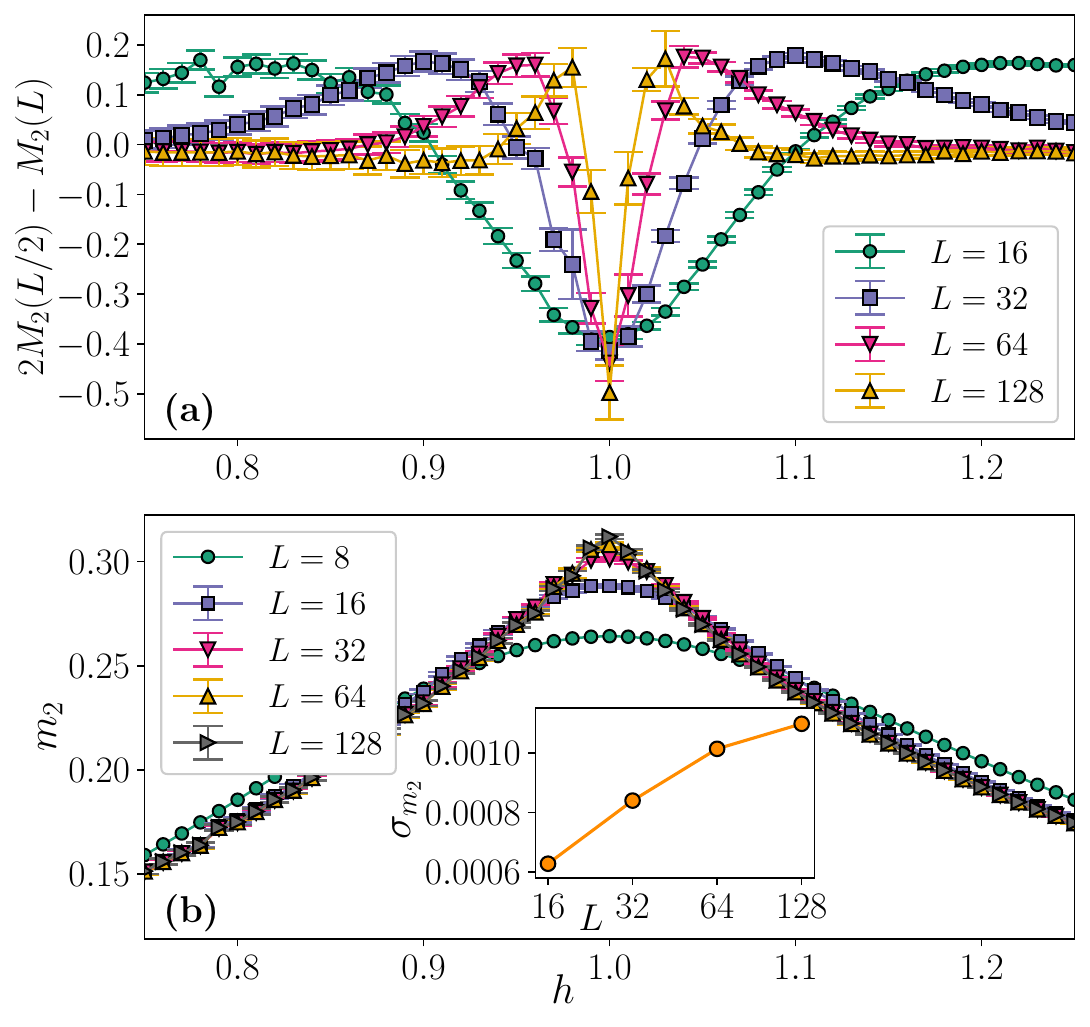}
    \caption{
    \textbf{Efficient estimation of Rényi-2 SRE density in 1D quantum Ising chain.}
    (a) The subleading term for the Rényi-2 SRE, $c_L = 2 M_2(L/2) - M_2(L)$, directly estimated using the efficient scheme specified in Sec.~\ref{sec:monte_carlo}, for various system-sizes in 1D quantum Ising chain.
    (b) The SRE density $m_2$ for the 1D quantum Ising chain near the critical point $h_c=1$ computed using the increment method using different subleading terms. (Inset) The sampling errors for $m_2$ at $h_c=1$ for various system-sizes $L$ (in log-scale). Clearly, the errors show even slower than 
    than logarithmic growth for the efficient sampling scheme. Here we consider TTN bond dimension $\chi=30$ and the number of sample is $N_S=10^6$. Error bars represent $95\%$ confidence interval.}
    \label{fig:ising_m2_subleading}
\end{figure}

The behavior of SREs in quantum Ising chain in 1D, i.e, 
\begin{equation}
    H_{\text{1D-Ising}} = -\sum_{\langle i,j \rangle} \sigma_i^x \sigma_{j}^x - h \sum_i \sigma^z_i,
    \label{eq:1dising}
\end{equation}
with $\sigma^{x, z}$ being the spin-1/2 Pauli matrices,
has been studied in Refs.~\cite{oliviero2022ising,haug2023quantifying}, where it has been shown that the SRE densities peak at the critical point $h_c=1$, and follow universal critical finite-size scaling hypothesis. In Fig.~\ref{fig:ising_m2_subleading}, we show the results for Rényi-2 SRE $M_2$, estimated efficiently using the subleading term $c_L = 2 M_2(L/2) - M_2(L)$ as described in Sec.~\ref{sec:monte_carlo}. Surprisingly, the sampling errors of the SRE density $m_2$ scales slower than $\log L$, with $L$ being the system-size, even at the critical point $h_c=1$. Therefore, 
unlike the MPS-based $2n$-replica method employed in Ref.~\cite{haug2023quantifying} that suffers from a computational cost of $O(\chi^{12})$, our Monte Carlo method for estimating $m_2$ provides accurate results without being severely limited by $\chi$.
Moreover, the computation of $m_2$ using the perfect sampling of MPS~\cite{lami2023quantum, haug2023stabilizer} will necessarily incur statistical errors that are exponential in system-size as the direct estimation of the subleading term $c_L = 2 M_2(L/2) - M_2(L)$ is not feasible by perfect sampling.

In the following, we extend the studies of SREs in 1D quantum many-body systems to qutrit systems by considering the three-state Clock model and the spin-1 XXZ model in 1D.

\subsubsection{Three-state Clock model} \label{sec:clock}

The quantum Clock model is a generalization of the quantum Ising model with $d$ states per site. Here we focus on the case $d=3$, where the Hamiltonian is
given by 
\begin{equation}
    H_{\text{1D-Clock}} =- \sum_{\langle i,j \rangle} (X_i X_j^\dagger + X_i^\dagger X_j) -h \sum_i (Z_i + Z_i^\dagger),
    \label{eq:clock}
\end{equation}
 where $X$, $Z$ are the shift and clock operators in Eq. \eqref{eq:clock} with $d=3$.
 The model is equivalent to the three-state Potts model \cite{Wu1982}. There is a transition from the ferromagnetic phase to the paramagnetic phase at $h_c=1$, as in the quantum Ising model. The critical point is described by $Z_3$ parafermion CFT, with central charge $c=4/5$. The exact correlation length exponent is $\nu_{\textrm{Potts}}=5/6$ \cite{Wu1982}.
It is to be noted that, since the system obeys $\Z_3$ symmetry, a two-site update scheme (see Sec.~\ref{sec:ttn_sampling}) is required to sample the Pauli strings that preserve the symmetry. Indeed, the Pauli strings that preserve the $\Z_3$ symmetry, generated by $\prod_i Z_i$, are generated by $Z_i$ and $X_i^\dagger X_j$ (up to a phase constant).

\begin{figure} 
    \centering
    \includegraphics[width=1\linewidth]{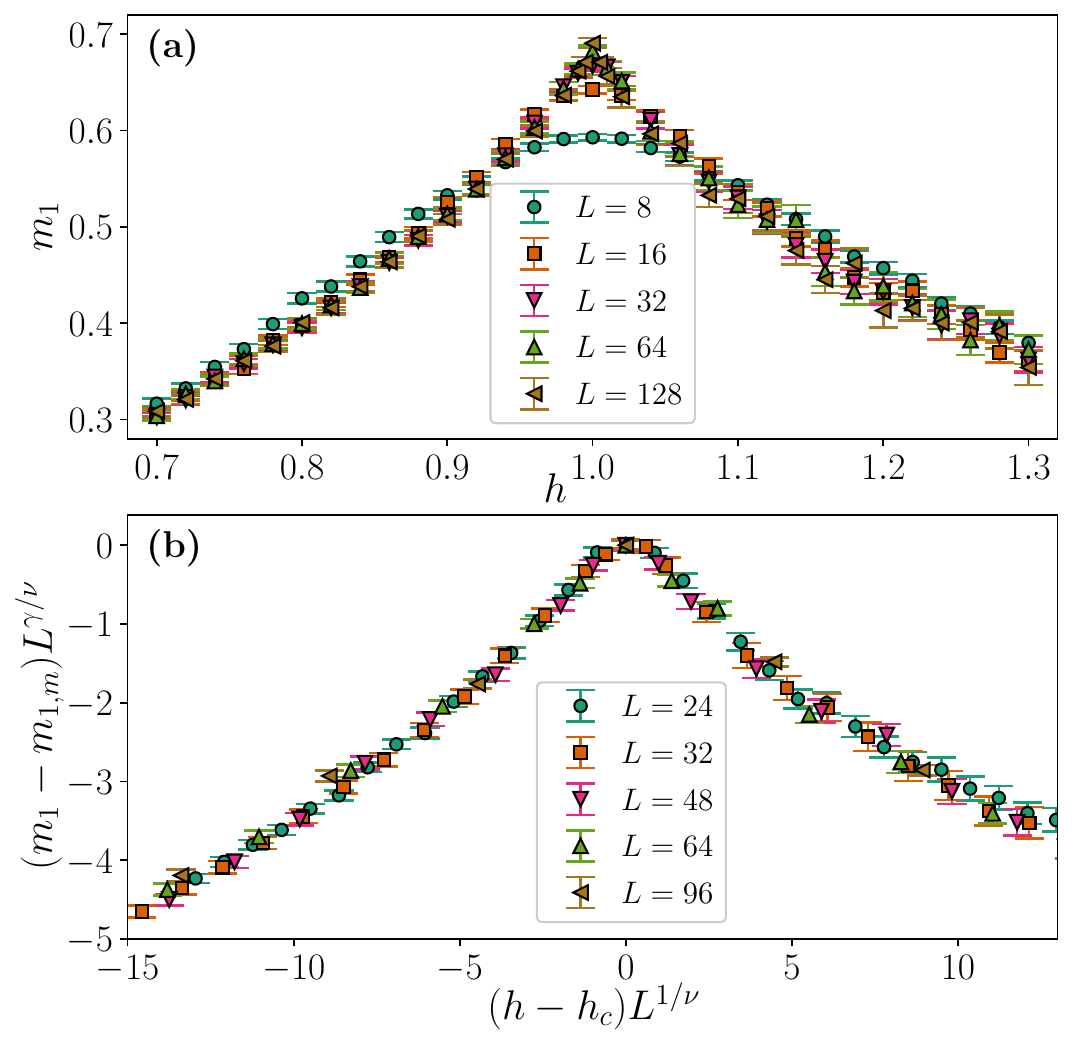}
    \caption{
    \textbf{Magic density in 1D quantum three-state Clock model.}
    (a) The SRE density $m_1$ in the ground-state of the three-state Clock model as a function of $h$. (b) Finite-size scaling for $m_1$. Here. $m_{1,m}$ is the maximum $m_1$ at $h_c=1$. We extract the critical exponent $\nu \approx 0.844$ and $\gamma \approx 0.66$. The correlation-length exponent $\nu$ is close to the known $\nu_{Potts}=5/6$. We used bond dimension up to $\chi=36$ and the number of sample is $N_S=10^6$. Error bars represent $95\%$ confidence interval.}
    \label{fig:magic_density_clock}
\end{figure}

In the three-state Clock model, the magic density displays similar behavior as in the quantum Ising model \cite{oliviero2022ising,haug2023quantifying}, as shown in Fig. \ref{fig:magic_density_clock}(a). Namely, $m_1$ displays maximum at the critical point $h_c=1$. We further investigate the finite-size scaling of $m_1$, that has been done for the quantum Ising chain \cite{haug2023quantifying}, using the finite-size scaling hypothesis:
\begin{equation}
    m_1 - m_{1, m}=L^{-\gamma/\nu}f \left( L^{1/\nu}(h-h_c) \right),
    \label{eq:fss}
\end{equation}
where $m_{1, m}$ is the maximum SRE density at $h_c=1$. 
In Fig. \ref{fig:magic_density_clock}(b), we show the data collapse corresponding to the finite-size scaling relation of Eq.~\eqref{eq:fss}, where we obtain the critical exponent $\nu \approx 0.844$, close to the expected theoretical value $\nu_{\textrm{Potts}}=5/6$.

\subsubsection{Spin-1 XXZ chain} \label{sec:xxz}

Next, we consider a spin-1 XXZ chain with single-ion anisotropy, whose Hamiltonian reads
\begin{equation}
    H_{\text{XXZ}} = -\sum_{\langle i,j \rangle} \left[ S^x_i S^x_j + S^y_i S^y_j + \Delta S^z_i S^z_j \right] + D \sum_{i} (S_i^z)^2,
\end{equation}
where $S^{\alpha}$'s, $\alpha = x, y, z$, are the spin-1 operators, $\Delta$ is the easy-axis anisotropy, and $D$ is the single-ion anisotropy. The model has a global $U(1)$ symmetry corresponding to the conservation of total magnetization $\sum_i S^z_i$, and here we consider the scenario of zero total magnetization. 

The phase diagram of the model has been studied in previous works \cite{chen2003,Tzeng2008, hu2011,Langari2013}. For $\Delta>0$, the model hosts three phases (with increasing $D$): the antiferromagnetic Néel order, the symmetry-protected topological (SPT) Haldane phase, and the large-$D$ trivial phase. The Néel to Haldane transition is an Ising transition, while the Haldane to large-$D$ transition is a Gaussian transition.

\begin{figure} 
    \centering
    \includegraphics[width=1\linewidth]{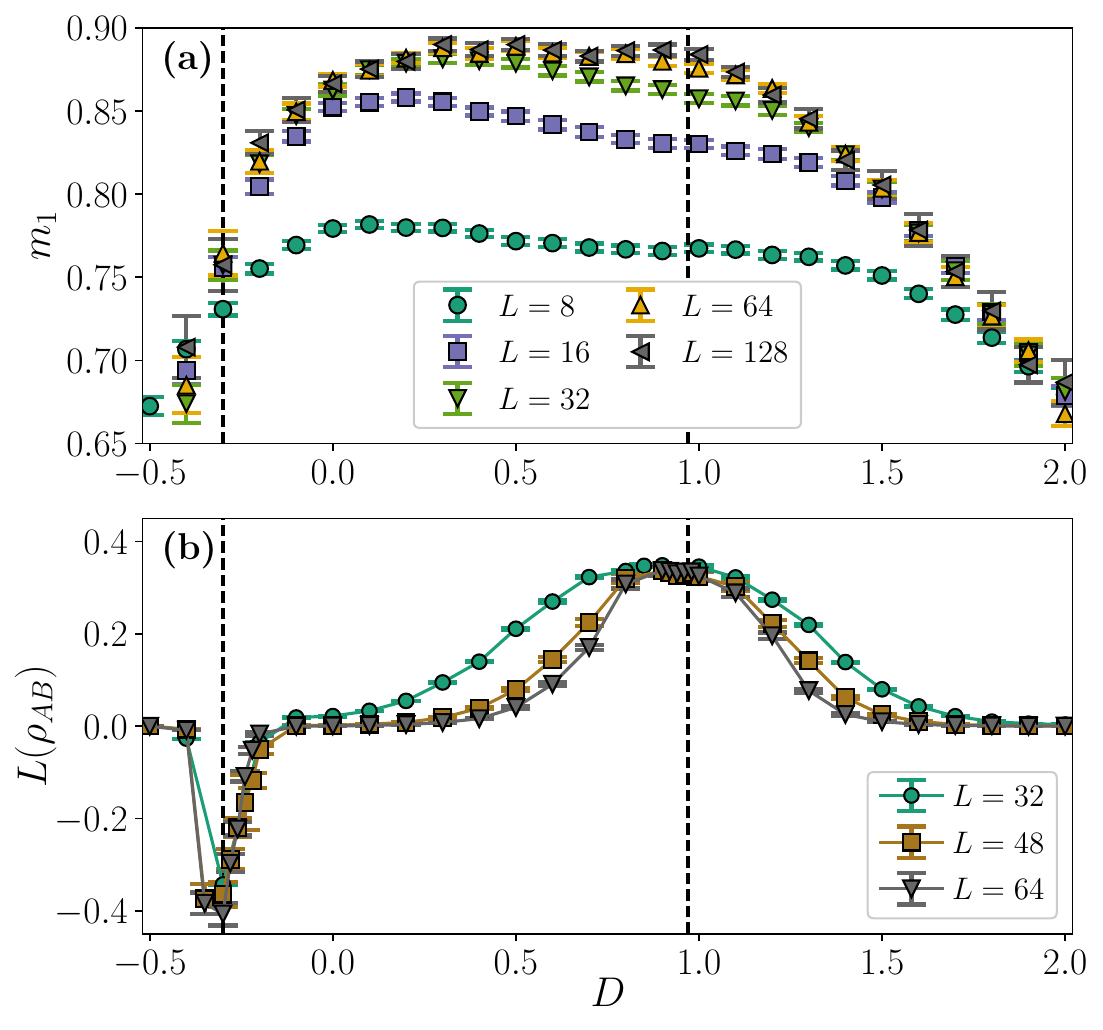}
    \caption{
    \textbf{Magic density and long-range magic in spin-1 XXZ chain.}
    (a) the magic density $m_1$ and (b) long-range magic $L(\rho_{AB})$ of the ground-state of the spin-1 XXZ model with $\Delta=1$ as a function of $D$.  We consider bond dimension up to $\chi=60$ and the number of sample is $N_S=10^6$. Error bars represent $95\%$ confidence interval. The dashed vertical lines represent the best estiamtes available for the transition points. }
    \label{fig:magic_density_xxz}
\end{figure}

Here, we focus on the isotropic case, i.e., $\Delta=1$. In this case, the transition is known to be at $D\sim -0.3$ and $D\sim 0.97$ for Néel-Haldane and Haldane-large $D$ transitions, respectively \cite{Tzeng2008, hu2011, Langari2013}. Fig.~\ref{fig:magic_density_xxz}(a) shows the SRE density $m_1$. We observe that $m_1$ is large and rather constant in the topological Haldane phase, while it becomes smaller in the neighboring phases. Note that the maximum value of $m_1$ for a product state is $\frac{2}{3} \log(4)\approx0.92$, achieved by the tensor product of single-qutrit states, each of which has $\langle P \rangle^2=1/4$ for all $P\neq I$. Thus, it is seen that the magic in the SPT Haldane phase almost saturates the maximum value.

\subsubsection{Long-range SRE}

In the spin-1 XXZ chain, while the onset of the topological Haldane phase is rather apparent from the magic density, there is no clear peak at the transitions, rendering the determination of the critical point difficult. Here we show that, unlike the magic density, the long-range magic $L_{AB}$ (see Eq.~\eqref{eq:lr_magic}), using the estimators in Eq. \eqref{eq:estimator_w} and Eq. \eqref{eq:estimator_i2}, can be used as a faithful indicator of quantum phase transitions. For the analysis of $L_{AB}$, we consider the spatially separated, extended subsystems $A=\{1,2,...,L/4  \}$ and $B=\{L/2+1,...,3L/4  \}$ in a perioidic chain of $L$ sites, as depicted in Fig. \ref{fig:schematic}b.

The long-range magic, for the the spin-1 XXZ chain, as plotted in Fig. \ref{fig:magic_density_xxz}(b) shows clear extremums at the two transitions. Although $L(\rho_{AB})$ is still non-zero for small $L$ away from criticality, it quickly decays to zero as the system size is increased. The peak at the Gaussian transition is very close to $D\sim 0.97$, as obtained with DMRG up to $L=20000$ spins~\cite{hu2011}. Notably, our results are obtained with only moderate sizes, and without any prior knowledge of the order parameter. At the Ising transition, the extremum occurs at a negative value as a minimum. Unlike entanglement, the SRE is not known to satisfy subadditivity, meaning that it is not always the case that $L(\rho_{AB}) \geq 0 $. Nevertheless, the non-trivial value at criticality is a useful indicator for detecting criticality.

The decay of long-range SRE away from criticality can be understood through a simple physical argument. Within a gapped phase characterized by a finite correlation length, when considering two subsystems $A$ and $B$ separated by a distance exceeding the correlation length, $A$ and $B$ are approximately uncorrelated. More formally, $\rho_{AB} \approx \rho_A \otimes \rho_B$, which implies $L(\rho_{AB})\approx 0$. In contrast, at criticality, the correlation length becomes infinite, such that $A$ and $B$ are always correlated regardless of their distance. This  results in a non-trivial value of $L(\rho_{AB})$.

\begin{figure} 
    \centering
    \includegraphics[width=1\linewidth]{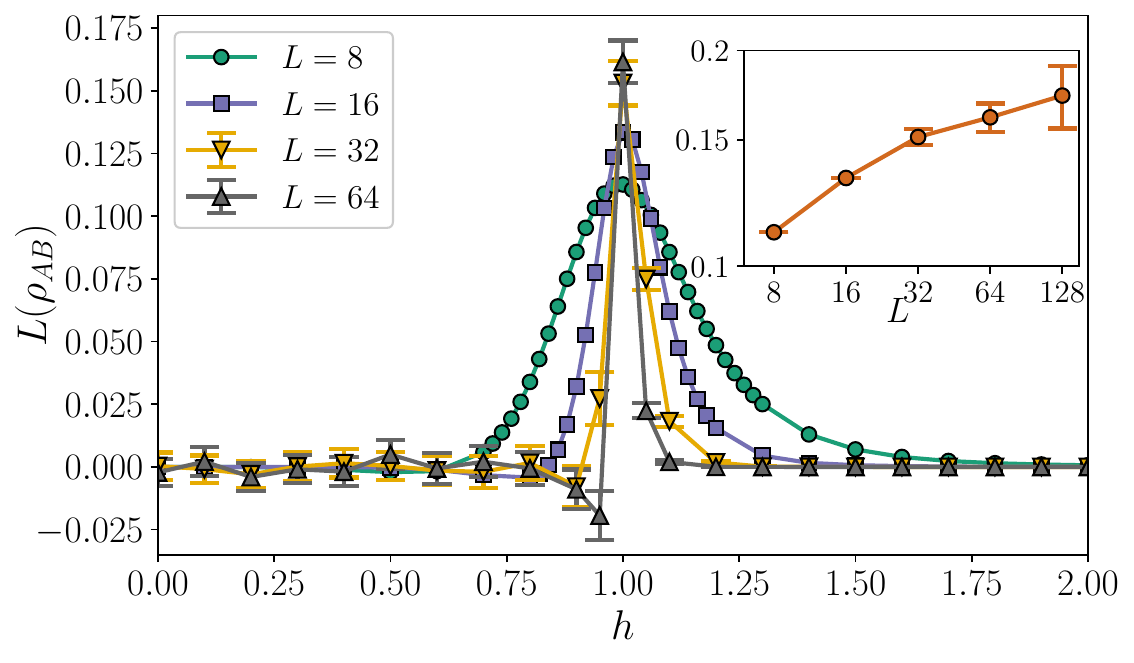}
    \caption{
    \textbf{Long-range magic in 1D quantum Ising chain.}
    The long-range magic $L(\rho_{AB})$ as in Eq. \eqref{eq:lr_magic} in the ground state of 1D quantum Ising chain as a function of the transverse field $h$. It peaks at the critical point $h_c=1$. (Inset) $L(\rho_{AB})$ at $h_c=1$ for various system sizes $L$ (in log-scale). 
    We consider TTN bond dimension up to $\chi=30$ and the number of sample is $N_S=10^6$.
    Error bars represent $95\%$ confidence interval.}
    \label{fig:lr_magic_ising}
\end{figure}

We also come back to the quantum Ising chain (Eq.~\eqref{eq:1dising}), and investigate the long-range magic across the Ising transition.
 We observe that $L(\rho_{AB})$ peaks at the critical point, as shown in Fig. \ref{fig:lr_magic_ising}. Furthermore, we plot $L(\rho_{AB})$ at $h_c=1$ in the inset of Fig. \ref{fig:lr_magic_ising}, where we see that the long-range magic grows logarithmically in $L$. In contrast, $L(\rho_{AB})$ quickly decays away from criticality (not shown). We note that, at the critical point, we observe long autocorrelation times between samples, which is the reason for the growing errors for larger sizes. This is reminiscent of the problem of critical slowing-down in the Monte Carlo simulations at criticality \cite{Wolff1990}. It is thus interesting to develop a cluster update, akin to Wolff cluster update \cite{Wolff1989}, that may overcome this issue, which we leave for future studies.

\subsection{SRE density in 2D many-body systems: $\Z_2$ lattice gauge theory} \label{sec:z2gauge}

Based on the favourable scaling of our scheme with system size, we investigate the non-stabilizerness in 2D systems, which so far have not been properly explored in the literature. In particular, we consider a $\Z_2$ lattice gauge theory, with Hamiltonian:
\begin{equation} \label{eq:z2}
    H_{\Z_2\text{-Gauge}} = -h\sum_\square \prod_{i\in \square} \tau^x_i - \sum_i \tau^z_i,
\end{equation} 
where the spin-1/2 Pauli operators, $\tau^{\alpha}$, $\alpha = x, z$, live on the links of the square lattice. The first term is the plaquette term that flips the four spins on an elementary square plaquette of the lattice. We are interested in the charge-free sector, that satisfies the Gauss' law
\begin{equation} \label{eq:gauss}
    \prod_{i\in +} \tau^z_i = 1,
\end{equation}
on each vertices of the lattice. It is well known that the Hamiltonian in Eq.~\eqref{eq:z2} is dual to the 2D transverse-field Ising model on the square lattice
\begin{equation} \label{eq:ising}
    H_{\text{2D-Ising}} =-\sum_{\langle i,j \rangle} \sigma^x_i \sigma^x_{j} -h \sum_i \sigma^z_i.
\end{equation}
by Wegner duality \cite{Wegner}. 
Here, the spin-1/2 Pauli operators, $\sigma^{\alpha}$, $\alpha=x, z$, live on the lattice sites of the dual square lattice.
It can be shown that the duality transformation preserves SREs (see Appendix \ref{sec:equivalence}). 
This enables us to compute the SREs of the the $\Z_2$ gauge theory \eqref{eq:z2} by considering the ground state of the transverse-field Ising model, which is computationally more convenient for TTNs. At the same time, our results also shed light on the transition point of the Ising model: there. the transition from ferromagnetic phase to the paramagnetic phase is known to be at $h_c\simeq 3.04$, as obtained with Quantum Monte Carlo \cite{Blote2002}. In the lattice gauge theory framework, such transition corresponds to confined to deconfined transition, where the behavior of Wilson loops turns from area to perimeter law. 

\begin{figure} 
    \centering
    \includegraphics[width=1\linewidth]{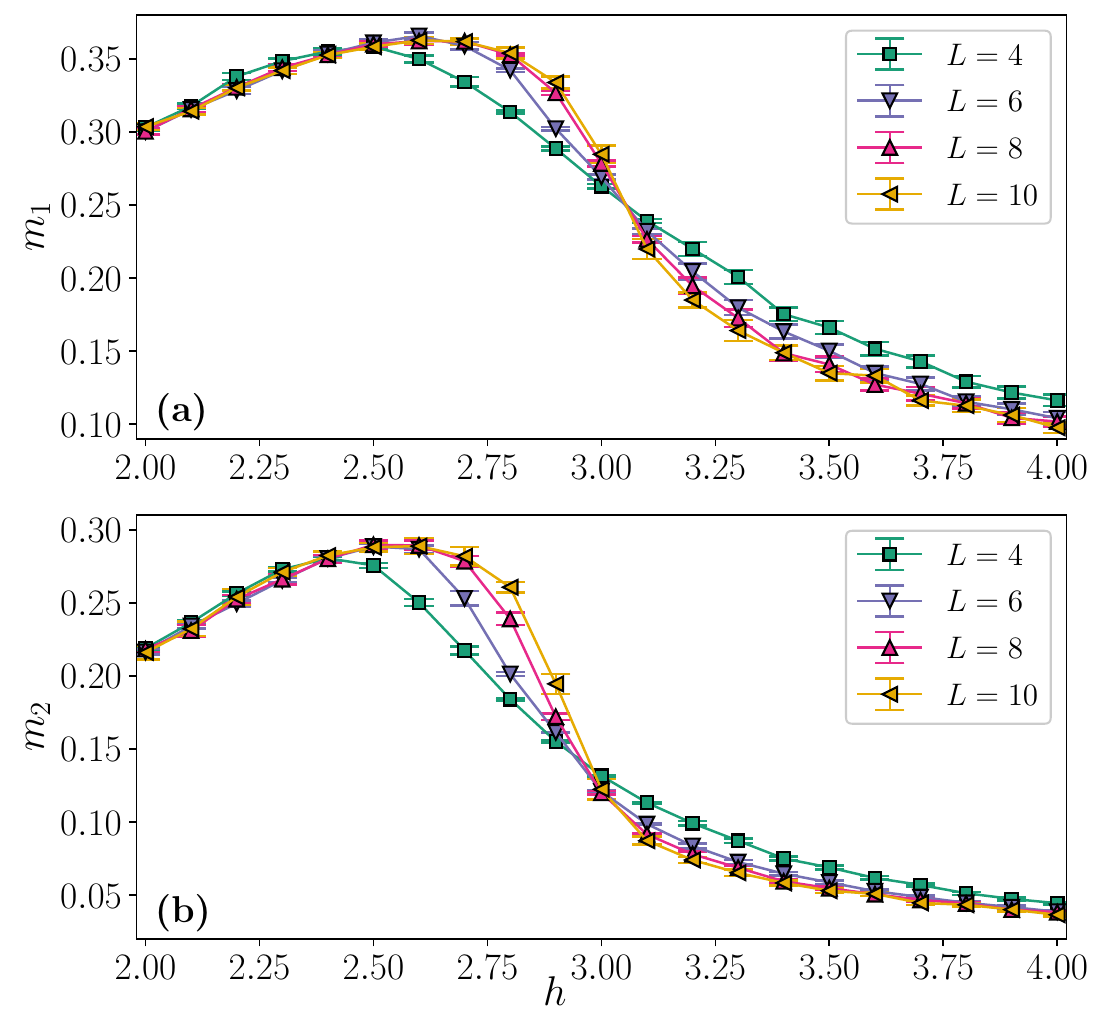}
    \caption{
    \textbf{Magic densities in 2D $\mathbb{Z}_2$ gauge theory.}
    The SRE desnities (a) $m_1$ and (b) $m_2$ of the ground-state of $\mathbb{Z}_2$ gauge theory on $L \times L$ square lattice as a function of $h$. We use TTN bond dimension up to $\chi=60$ and the number of sample is $N_S=10^6$. Error bars represent $95\%$ confidence interval.}
    \label{fig:magic_density_ising}
\end{figure}

The results for magic density for $n=1,2$ are presented in Fig.~\ref{fig:magic_density_ising}. It is seen that both quantities detect the transition. However, the observed behavior is very different from the 1D quantum Ising chain, which exhibits a peak at the transition. Instead, here we observe that the curves exhibit crossings at the transition. 

\begin{figure} 
    \centering
    \includegraphics[width=1\linewidth]{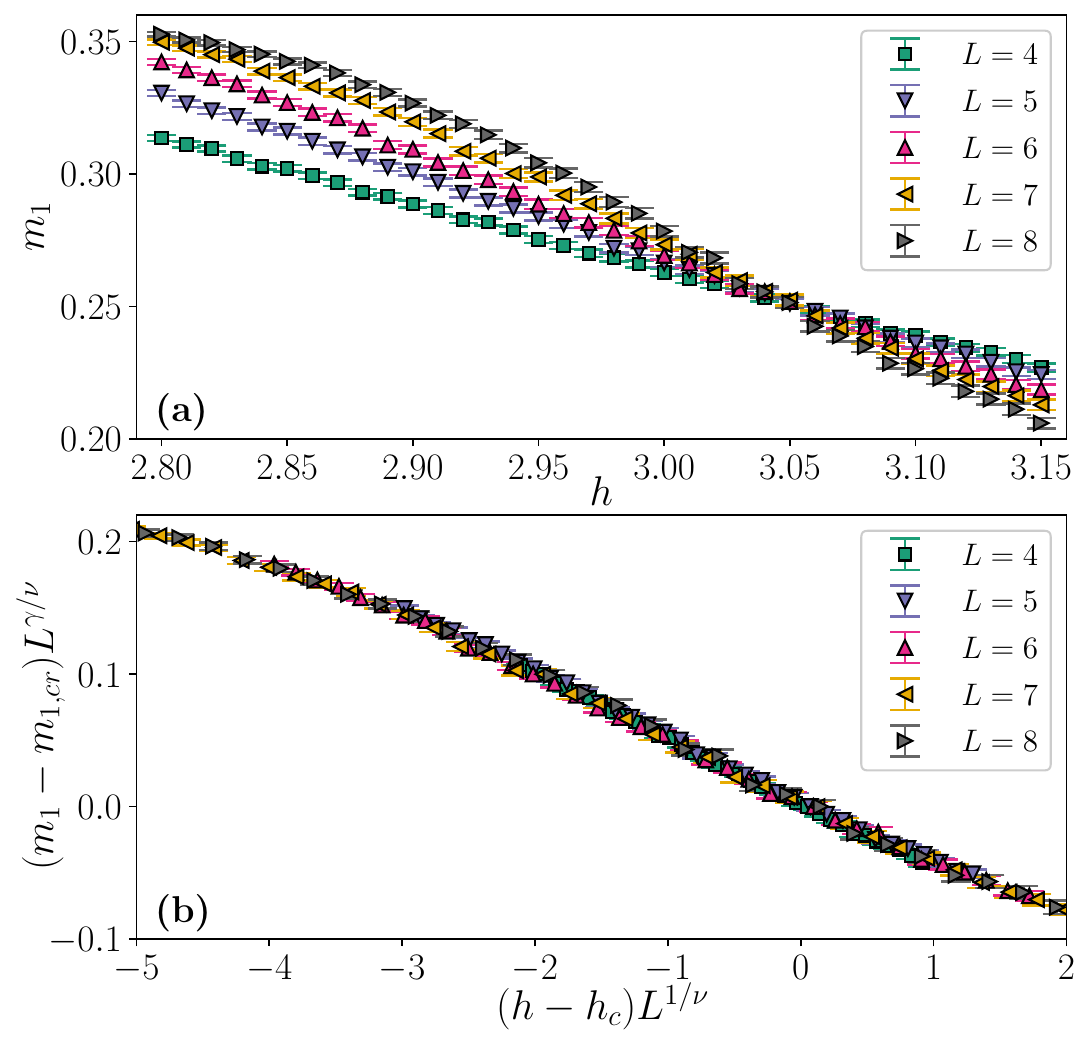}
    \caption{
    \textbf{Finite-size critical scaling of SRE density in 2D $\mathbb{Z}_2$ gauge theory.}
    (a) The SRE density $m_1$ near the critical point at the $\mathbb{Z}_2$ gauge theory. Even with small TTN bond dimension $\chi=30$, $m_1$ captures the transition very well: all the curves cross near the known critical point $h_c=3.04$. (b) Finite-size scaling of $m_1$. Here, $m_{1,cr}$ is $m_1$ at $h=3.04$. We find the correlation length critical exponent $\nu=0.64 \pm 0.05$. The extracted $\nu$ is remarkably close to the known $\nu_{3D}\simeq 0.63$ for 3D Ising universality class. Here, the number of samples is $N_S=10^7$.}
    \label{fig:magic_crit}
\end{figure}

In Fig.~\ref{fig:magic_crit}(a), we depict $m_1$ close to the critical point,
using a fixed bond dimension $\chi=30$. Remarkably, we observe that $m_1$ detects the transition point very well: all the curves cross near the critical point at $h_c=3.04(1)$. We should highlight at this point that the TTN ansatz with such a low bond dimension of $\chi=30$ can not approximate the ground state wave function accurately near the critical point, particularly in 2D critical systems.
Consequently, the standard phase transition detectors, such as the Binder cumulant, calculated from the TTN state with $\chi=30$, do not exhibit the expected critical crossing behavior -- see Appendix~\ref{sec:binderIsing2D} for a direct comparison in the present case.
Therefore, 
the remarkable observation of the perfectly crossing behavior in $m_1$ near the critical point underscores the significant value of magic in detecting and characterizing quantum phase transitions. This is particularly relevant in situations where other quantities are prone to significant errors, e.g., due to limited bond dimensions in tensor network states. While we believe that a further characterization of what the scaling resources (e.g., size and bond dimension) to detect a transition point are
is outside the scope of our paper, this would be very much worth pursuing based on the Ising model results we presented.

Furthermore, we show excellent data collapse for $m_1$ in Fig. \ref{fig:magic_crit}(b), using the finite-size scaling relation of Eq.~\eqref{eq:fss}, from which we extract the correlation length exponent $\nu=0.64 \pm0.05$, that is close to the known $\nu_{3D}=0.63$ for 3D (classical) Ising universality~\cite{Blote2002}.

\section{Experimental protocol}
\label{sec:exp_protocol}

The numerical method described above can be easily adapted for experimental measurements of SREs. In particular, we can sample Pauli strings according to $\Xi_P$ using Monte Carlo sampling. We note that, although the probability distribution $\Xi_P$ can be sampled directly through measurements in the Bell basis \cite{montanaro2017,haug2023}, the method requires preparation of two copies of a state and joint operations on them. In practice, this may not be feasible in some experimental platforms, or difficult to scale up to larger sizes and higher-dimensional systems. Moreover, the method only works for real wavefunctions \cite{Gross2021}. Instead, our proposal relies solely on measurements in the computational basis on a single instance of a state, and it is applicable to generic quantum states.

In experiments, the Pauli strings are measured from $N_M$ copies of $\rho$ where the  measurement outcomes are $A_i\in \{ +1,-1\}$. The expectation value is then given by the average taken over the random measurement outcomes. The sampling of Pauli strings can be performed with Metropolis algorithm, similar to our numerical calculations. 
However, it is important to note that in experimental setups, the candidate Pauli string is not restricted to few-site updates, as is the case of TTN. This flexibility allows for multi-site updates and can potentially reduce the autocorrelation time associated with the sampling process enormously.

For a finite number of measurements $N_M$, we have that 
\be \label{eq:estimator_p}
    \bar{P} = \frac{1}{N_M} \sum_{i=1}^{N_M} A_i
\ee
is an estimate for $\langle P \rangle$. The total number of resources is thus $N_M \times N_S$, where $N_S$ is the number of sampled Pauli strings. In view of Eq. \eqref{eq:var}, when the SREs are at most $O(\log N)$, the required $N_S$ is polynomial in $N$. Note that $N_M$ may still be exponential, but it is expected to be no larger than $O(d^N)$, with $d$ being the local dimension. As a result, the number of resource required in our protocol is significantly lower than the protocol in Ref. \cite{leone2023} when the SREs are at most $O(\log N)$. Moreover, our protocol offers a possibility to measure $M_1$, in which case $N_S$ is always polynomial\footnote{
It is to be noted that the measurement of $M_n$ with $n \neq 1$ in experiments by employing the increment scheme with subleading terms, as discussed in Sec.~\ref{sec:monte_carlo}, can be  challenging. This procedure necessitates the simultaneous sampling from two distinct physical systems, something that is easily achievable on some platforms (optical lattices, circuit QED, tweezer arrays) but not immediately on others (e.g., ion chains). Furthermore, in experimental measurements, the obtained expectation values are only approximations of the true values. Consequently, computing ratios of these approximate values, as in Eq.~\eqref{eq:subleading}, introduces errors into the calculations.}.

The variance of the estimator in Eq. \eqref{eq:estimator_p} is given by $\textrm{Var} (P)=1-\langle P \rangle^2$. Thus, the standard error reads

\be
    \Delta P = \sqrt{ \frac{1-\langle P \rangle^2}{N_M}}.
\ee
For large $N_M$, the random variable $\bar{P}$ approximately has a Gaussian distribution with average $\langle P \rangle$ and standard deviation $\Delta P$. Note that this will introduce bias to the estimators in Eq. \eqref{eq:estimator_n} and Eq. \eqref{eq:estimator_1}. This bias can be made smaller by increasing $N_M$, where the estimators become unbiased in the limit $N_M \rightarrow \infty$.   

\begin{figure}  
    \centering
    \includegraphics[width=1\linewidth]{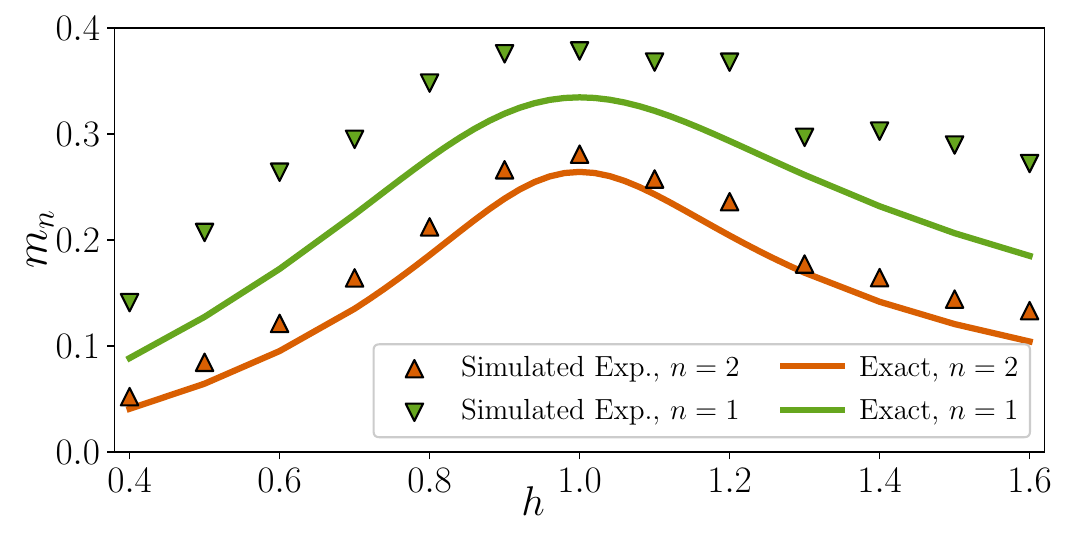}
    \caption{
    \textbf{Simulated experiment to measure SREs.}
    Simulation of experimental measurement of SREs in the ground state of 1D quantum Ising chain for $L=8$. Here, $N_M=500$ and $N_S=10^4$.}
    \label{fig:exp_ising}
\end{figure}

Here, we simulate this situation numerically by perturbing the computed $\langle P \rangle$ with $\epsilon$, where $\epsilon$ is a random number chosen from a Gaussian distribution centered at zero and with standard deviation $\Delta P$. We would like to investigate the effects of taking finite $N_M$ and $N_S$. Here, we consider the ground state of 1D transverse-field Ising chain at $h=1$ for concreteness. An example of the results of such a protocol is shown in Fig. \ref{fig:exp_ising} for $L=8$ with $N_M=500$ and $N_S=10000$.

\begin{figure}  
    \centering
    \includegraphics[width=1\linewidth]{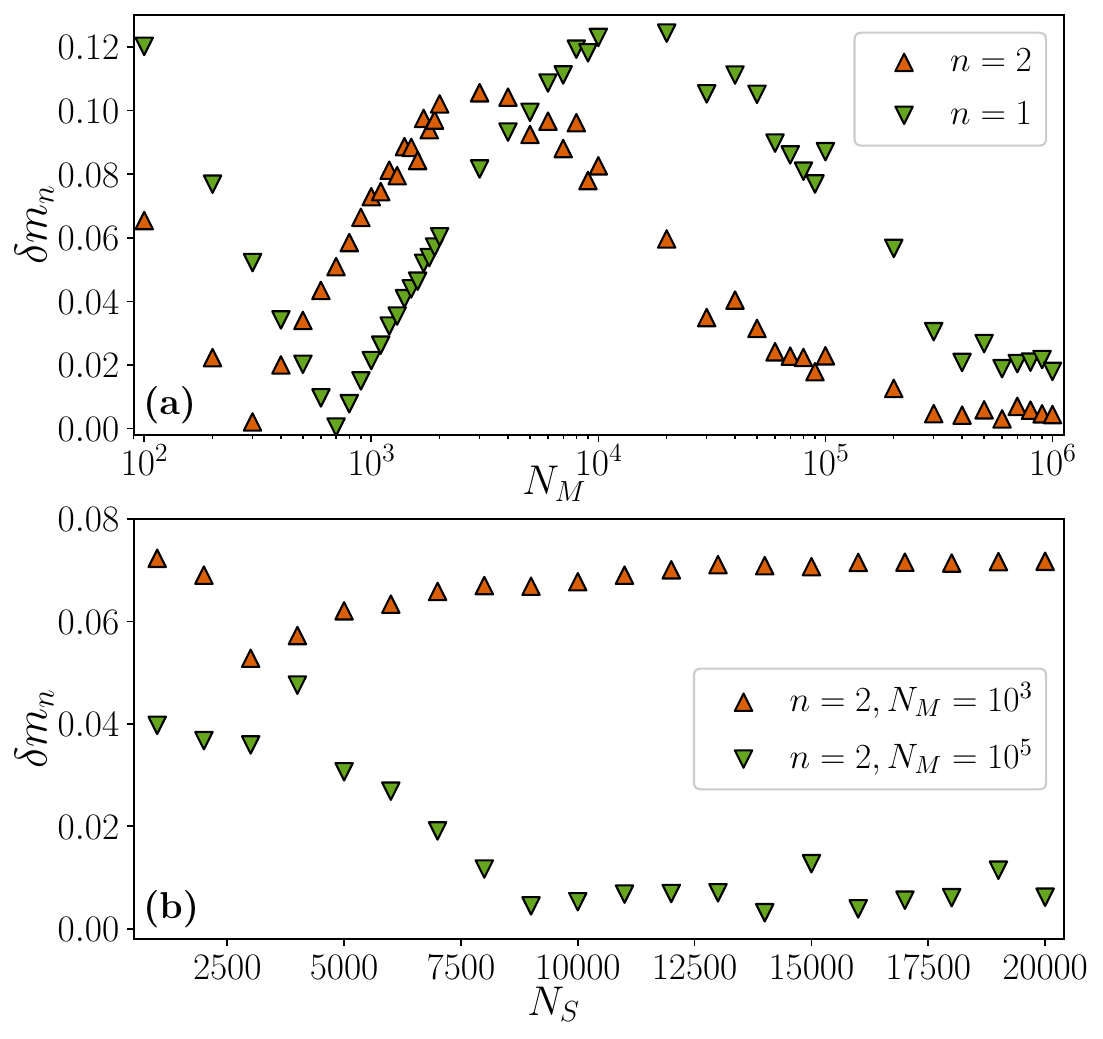}
    \caption{
    \textbf{The errors in SRE density for simulated experiments.}
    The deviation $\delta m_n = |m_{n, \text{Sim.Exp.}} - m_{n, \text{exact}}|$ for $n=1,2$ in the ground state of 1D quantum Ising chain  at the critical point $h=1$ for $L=16$. In (a), we fix $N_S=10000$ and vary $N_M$, while in (b), we fix $N_M=\{10^3,10^5\}$ and vary $N_S$.}
    \label{fig:exp_error}
\end{figure}

Next, we compute the deviation $\delta m_n = |m_{n,\text{Sim.Exp.}}-m_{n,\textrm{exact}}|$ for $n=1,2$, where
$m_{n,\text{Sim.Exp.}}$ denotes the SRE density in simulated experiments.
The results are shown in Fig. \ref{fig:exp_error}. We see that, for fixed $N_S$, the error first increases for small $N_M$, before it eventually decreases. We expect this is due to the bias with finite number of Pauli measurements, as mentioned above. Indeed, as shown in Fig. \ref{fig:exp_error}(b), we see that increasing $N_S$ while fixing $N_M$ does not result in vanishing $\delta m_n$.

\section{Conclusions and outlook} \label{sec:concl}
We have proposed a Markov chain Monte Carlo approach to compute magic in many-body systems. We have discussed how the full state magic $M_n$ can be estimated for different values of $n$, and demonstrated the corresponding efficiency in several scenarios. Moreover, long-range magic can be estimated efficiently in general. The implementation of our algorithm is flexible and compatible with various wave-function based methods. Specifically, we have provided detailed insights into the efficiency and flexibility of our method when applied to tree tensor networks.

Through our algorithm's flexibility, we have gained valuable insights into the role of magic in many-body systems. In one-dimensional systems, we observed that full state magic is not universally associated with critical behavior. While it displays criticality signatures in certain cases like Ising and Potts models, it does not in others. However, long-range magic overcomes this limitation and consistently exhibits indications of critical behavior across all scenarios we investigated. We speculate that the functional form of long-range magic, similar to mutual information, is free of potential UV-divergences in a field theory framework.

The very mild volume scaling cost of our sampling has also enabled us the exploration of two-dimensional $\mathbb{Z}_2$ lattice gauge theories. There, we have found that magic displays finite-volume crossings in correspondence of the confined-deconfined phase transition, and it also follows universal scaling behavior up to the volumes (100 spins) we were able to treat. Remarkably, magic was well converged even at modest bond dimensions.

Our numerical results suggest a deep connection between (long-range) magic and many-body properties, highlighting the direct links between stabilizer Renyi entropies and physical phenomena such as quantum critical behavior and confinement-deconfinement transitions. To complement our theoretical findings, we have proposed an experimental protocol for measuring stabilizer Renyi entropies solely using measurements in the computational basis.

In terms of future investigations, our technique can be extended to explore nonstabilizerness in finite-temperature scenarios by generalizing it to tree-tensor operators that efficiently represent low-temperature many-body states. In particular, it would be interesting to study the behavior of stabilizer Renyi entropies at finite-temperature phase-transition and compare it with other information-theoretic quantities, such as entanglement \cite{Lu2020,Lu_2019_2,Wald_2020,Arceci2022}, quantum discord \cite{Tarabunga2022}, and quantum coherence \cite{frerot2022}. Along the same lines, another possible scenario would be applying our tools to faulty quantum circuits, recently discussed in the context of magic in Ref.~\cite{niroula2023phase}. It would also be instructive to perform a systematic investigation of magic within topological phases, extending our analysis of the Haldane phase. 
Another interesting perspective is to understand the role of magic in many-body quantum dynamics of closed quantum systems, whose investigation in the context of Ising models has been the subject of recent works \cite{rattacaso2023stabilizer}. In particular, our method allows for the investigation of genuine long-distance magic, that might be instrumental in establishing the presence or absence of propagation bounds for magic.

At the methodological level, our work opens a series of questions. The Markov chain Monte carlo approach could be extended to investigate other magic measures that depend only on expectation values, such as mana. Moreover, so far, we have only employed very basic sampling strategies. It would be worth exploring how different ones, such as heatbath or non-local updates, can be used to design better magic estimators since, in terms of experimental applicability, having shorter autocorrelations could considerably improve realistic implementations. In terms of efficiency of the increment trick in 2D models, it would be interesting to study whether a one-dimensional projection of 2D systems such as the one introduced in \cite{kadowasa2023} would be beneficial. Finally, it would be interesting to understand the finer structure of sampling Pauli strings in many-body systems, that could reveal both useful insights into novel algorithms, and potentially deeper connections between many-body properties and magic.

\acknowledgements
We are indebted to M. Collura, A. Hamma, G. Lami, L. Leone, and S. F. E. Oliviero for fruitful discussions and collaborations on related topics. We thank G. Magnifico, S. Montangero, S. Notarnicola and P. Silvi for discussions on tree tensor networks. We thank L. Piroli and C. Castelnovo for comments on the manuscript. 
P.S.T. acknowledges support from the Simons Foundation through Award 284558FY19 to the ICTP. 
M.D. and E.T. acknowledge support from the MIUR Programme FARE (MEPH), and from  QUANTERA DYNAMITE PCI2022-132919. M.D. work was also supported by the PNRR MUR project PE0000023-NQSTI, and by the EU-Flagship programme Pasquans2. E.T. is also funded by the European Union under Horizon Europe Programme - Grant Agreement 101080086 — NeQST.
Our TTN codes have been implemented using C++ Itensor library~\cite{itensor22}.

\appendix

\section{Autocorrelations and statistical errors}
\label{sec:auto_corr}

\begin{figure}
    \centering
    \includegraphics[width=1\linewidth]{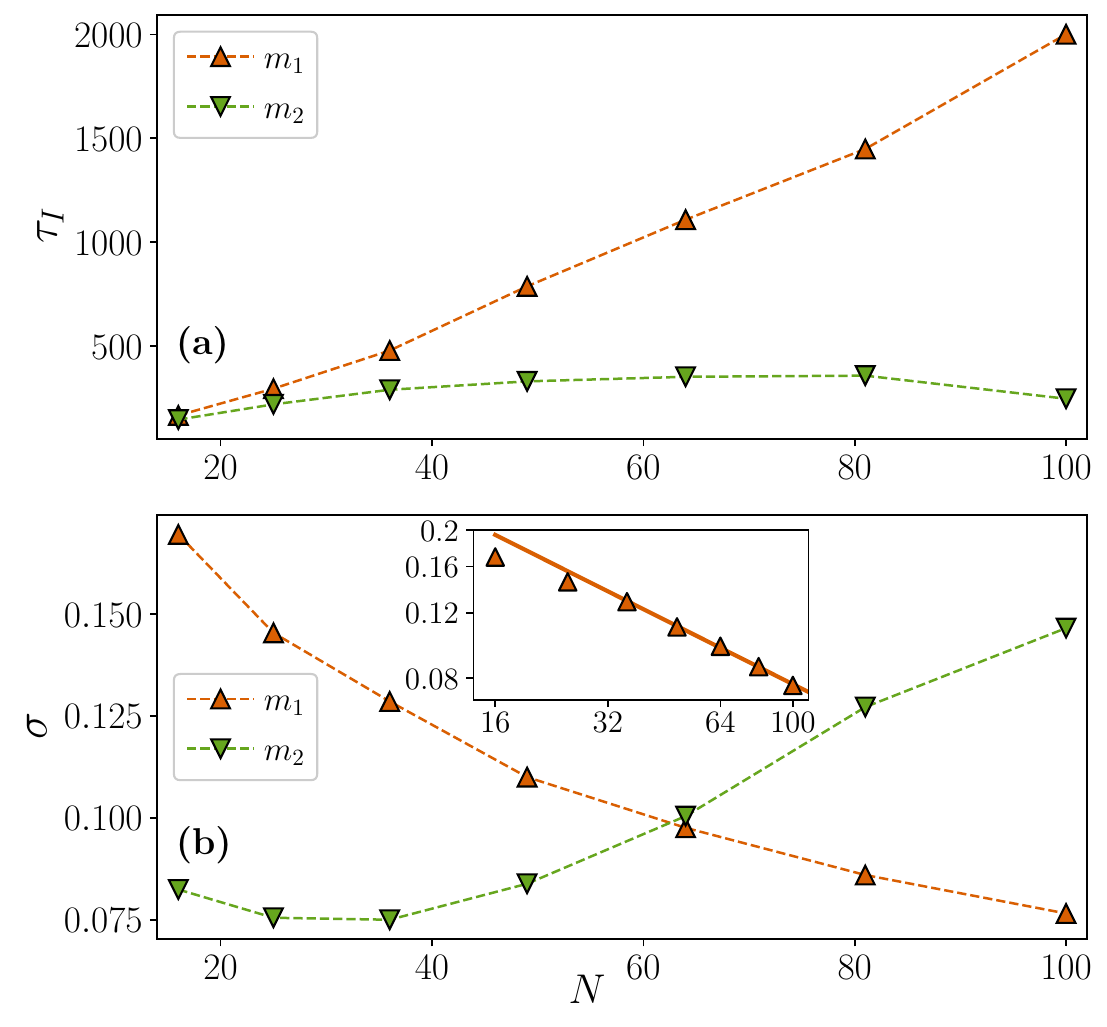}
    \caption{
    \textbf{Autocorrelation time and statistical errors in Monte Carlo sampling of SREs.}
    (a) Integrated autocorrelation time $\tau_I$ at the ground state of 2D transverse-field Ising model with $h=3$ for various system sizes $N=L \times L$. It is linear for $m_1$ and saturates for $m_2$. (b) Standard deviation $\sigma$ for various system sizes. Inset shows $\sigma$ for $m_1$ in log-log scale. The solid line denotes a fit $\sigma=aN^{-b}$ for $L \geq 6$, with $b=0.503$. The standard deviation is obtained by error propagation.}
    \label{fig:auto_time}
\end{figure}

Here, we analyzed the integrated autocorrelation time of $m_1$ and $m_2$ close to the critical point of the 2D transverse-field Ising model at $h=3$. The integrated autocorrelation time is defined as $\tau_I = 1+ 2\sum_{t=1}^\infty \rho(t)$, where $\rho(t)$ is the autocorrelation function. 
The integrated autocorrelation time affects the statistical errors of the averages obtained from Monte Carlo sampling \cite{Sandvik2010}. We observe that $\tau_I$ is linear for $M_1$, while it saturates for $M_2$, as shown in Fig. \ref{fig:auto_time}(a). We have also checked that $\tau_I$ does not show much variation with respect to bond dimension.

Moreover, the standard deviation $\sigma$ for various system sizes is shown in Fig.~\ref{fig:auto_time}(b). For $n=2$, it is seen that $\sigma$ grows exponentially, confirming the analysis in Sec.~\ref{sec:monte_carlo}. On the other hand, for $n=1$, $\sigma$ is decreasing with power-law behavior (see inset). The power-law exponent is found to be compatible with $1/2$, again as anticipated in Sec.~\ref{sec:monte_carlo}.
We note here that the behavior of the integrated autocorrelation time and the standard deviations remains qualitatively similar near the critical points for other many-body systems considered here.

\section{Convergence with bond dimension} \label{sec:bond_dim}
\begin{figure}
    \centering
    \includegraphics[width=1\linewidth]{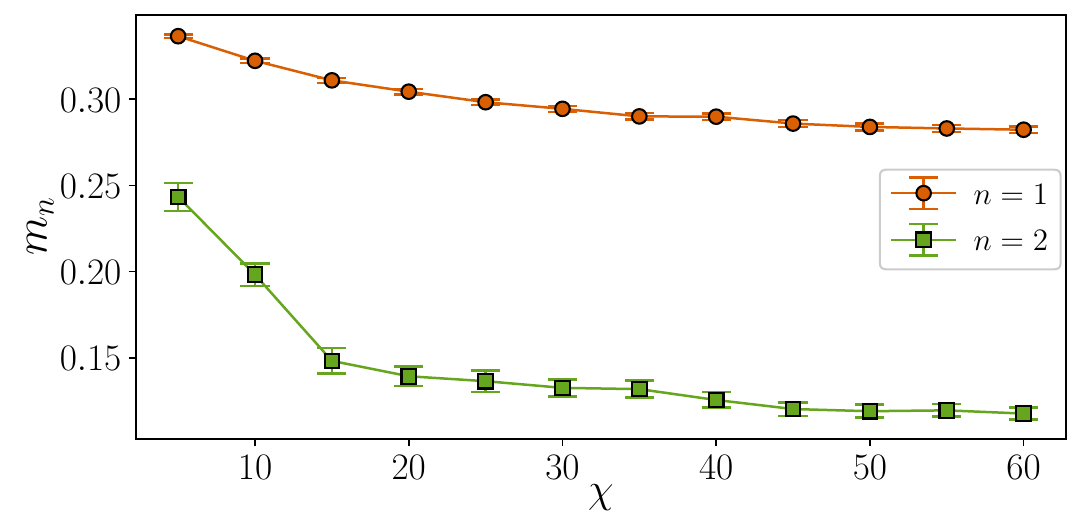}
    \caption{
    \textbf{Convergence of SRE with respect to bond dimension.}
     SREs $m_1$ and $m_2$ at the ground state of 2D transverse-field Ising model with $h=3$ and $L=10$ for various bond dimension $\chi$.}
    \label{fig:bond_dim}
\end{figure}

In our simulations, we have ensured that the SREs have converged with bond dimensions of the TTN in each models. To this end, we carried out simulations with different bond dimensions and verified that the SREs have sufficiently converged within statistical accuracy, which is typically on the order of $10^{-3}$.
Fig. \ref{fig:bond_dim} illustrates an example of the dependence of the SREs $m_1$ and $m_2$ for the ground state of the 2D transverse-field Ising model with linear size $L=10$. We see that as the bond dimension $\chi$ is increased, the SREs eventually converge to a constant within error bars.

\section{Equivalence between 2D $\Z_2$ gauge theory and the 2D transverse-field Ising model}
\label{sec:equivalence}

The duality transformation between Eq. \eqref{eq:z2} and Eq. \eqref{eq:ising} is defined with the following transformation,
\begin{equation} \label{eq:ising_var}
\begin{split}
    \sigma^x_i \sigma^x_j &= \tau^z_{\langle ij \rangle} \\
    \sigma^z_i &= \prod_{i\in \square} \tau^x_i. 
\end{split}
\end{equation}
 More precisely, the transformation maps the charge-free sector of Eq. \eqref{eq:z2} to the even sector of Eq. \eqref{eq:ising}.

It is easy to see that the mapping in Eq. \eqref{eq:ising_var} maps Pauli strings in the Ising model to Pauli strings in $\Z_2$ gauge theory, because the Pauli operators on both sides of the equation generate the Pauli group in the corresponding models. Since the SREs depend only on the expectation values of Pauli strings, it follows that the SREs are preserved by the duality transformation. Therefore, the SREs in the Ising model are identical to the SREs in $\Z_2$ gauge theory. It should be, however, noted that 
equivalence relation in case of
the subsystem mixed-state SRE (e.g., $\tilde{M}_2$ defined in Eq.~\eqref{eq:sre_mixed}), and the long-range magic thereof, is non-trivial because of the non-local nature of the transformation \eqref{eq:ising_var}. Consequently, the distribution of magic within the subsystems may differ in these two theories.

It is worth nothing that the same conclusion evidently holds for other dualities that map Pauli strings to Pauli strings, such as the Kramers-Wannier duality which maps $h \to h^{-1}$ in Eq. \eqref{eq:1dising} and Eq. \eqref{eq:clock}. As previously discussed, the long-range magic is not preserved under the duality. This is reflected in the distinct behavior of $L(\rho_{AB})$ for $h>1$ and $h<1$ in Fig. \ref{fig:lr_magic_ising}.

\section{Binder cumulant in 2D quantum Ising model with TTN}
\label{sec:binderIsing2D}

\begin{figure}
    \centering
    \includegraphics[width=\linewidth]{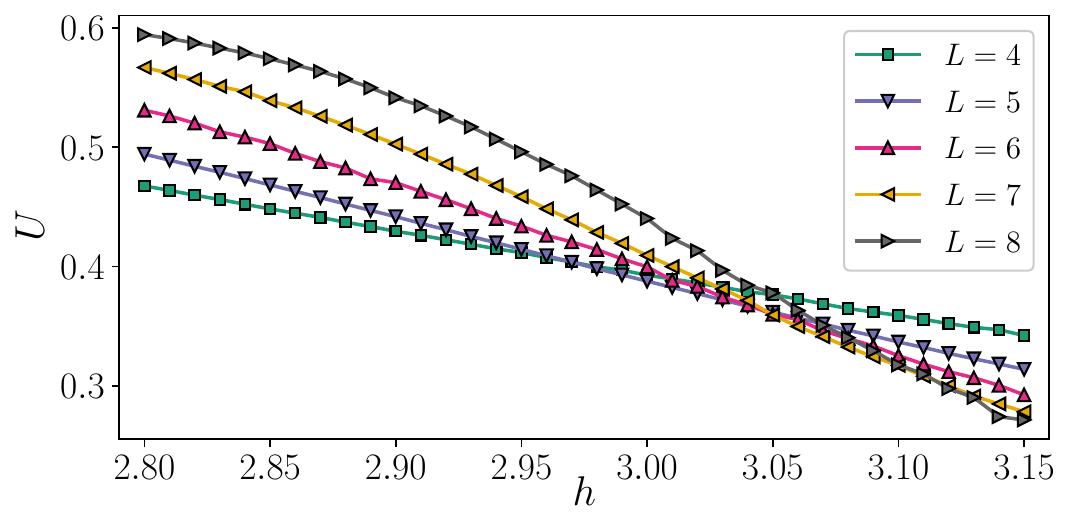}
    \caption{
    \textbf{The Binder cumulant across the critical point in the 2D quantum Ising model.} Here we approximate the ground state of 2D quantum Ising model with TTN having bond dimension $\chi=30$, in parity with Fig.~\ref{fig:magic_crit}.
    }
    \label{fig:binder_2dising}
\end{figure}

In Sec.~\ref{sec:z2gauge}, we have demonstrated the ability of the magic density to accurately detect and characterize the quantum critical point in the 2D $\mathbb{Z}_2$ gauge theory, and thereby in 2D quantum Ising model.
Notably, the curves of $m_1$ for different linear system-sizes exhibit a clear critical crossing behavior near the critical point $h_c=3.04$, even with a modest TTN bond dimension of $\chi=30$. However, the same level of accuracy is not achieved when utilizing the Binder cumulant, defined as
\begin{equation}
    U = 1 - \frac{\langle s_x^4 \rangle}{3 \langle s_x^2 \rangle^2}, \quad \text{with } s_x = \frac{1}{L^2}\sum_i\sigma^x_i,
\end{equation}
for the 2D Ising model \eqref{eq:ising}.
Due to the inability of the TTN state with a small bond dimension of $\chi=30$ to faithfully represent the ground state in the vicinity of the critical point, the calculation of the Binder cumulant $U$ yields erroneous results. Consequently, the curves of $U$ for different linear system-sizes $L$ do not exhibit a clear crossing behavior near the critical point (Fig.~\ref{fig:binder_2dising}). For instance, while the curves for $L=4$ and $5$ intersect at $h=2.98$, the intersection for $L=7$ and $8$ occurs around $h=3.14$. As such, if one attempts to perform finite size scaling on the Binder cumulant data, the resulting critical point and the correlation-length critical exponent $\nu$ will be erroneous.

%\addcontentsline{toc}{chapter}{Bibliography} 
\bibliography{biblio.bib}

\end{document}